\documentclass[12pt]{article}
\usepackage{amsmath}
\usepackage{amsfonts}
\usepackage{amssymb}
\usepackage{amsthm}
\usepackage[mathscr]{eucal}
\usepackage[mathscr]{eucal}

\usepackage[nosort]{cite}

\def\theequation{\arabic{section}.\arabic{equation}}
\begin{document}

\vspace*{0.4cm}


\begin{center}

{\LARGE\bf  Two-twistor particle models  }

\vspace{0.2cm}

{\LARGE\bf and free massive higher spin fields}

\vspace{1.4cm}

{\large\bf J.A. de Azc\'arraga}$^{\,a}$,\,\,\, {\large\bf S. Fedoruk}$^{b\,c\,\ddagger}$
,\,\,\,
{\large\bf J.M. Izquierdo}$^{\,d}$,\,\,\, {\large\bf J. Lukierski}$^{\,e}$
\begin{center}
\center{
{\qquad}\item[$^a$]
{\it Dept. Theor. Phys. and IFIC (CSIC-UVEG), Valencia Univ. \\
46100-Burjassot (Valencia), Spain}\\
 {\tt j.a.de.azcarraga@ific.uv.es}
   {\qquad}\item[$^b$]
{\it Bogolubov Laboratory of Theor. Physics, JINR, \\  Joliot-Curie 6,
141980 Dubna, Moscow region, Russia}\\
{\tt fedoruk@theor.jinr.ru}
   {\qquad}\item[$^c$]
{\it Dept. Theor. Phys., Tomsk State Pedagogical University, \\
Tomsk, 634061, Russia}
 {\qquad}\item[$^d$]
{\it Dept. Theor. Phys., Valladolid Univ., 47011-Valladolid, Spain}\\
{\tt izquierd@fta.uva.es}
{\qquad}\item[$^e$]
{\it Institute of Theoretical Physics, Wroc{\l}aw Univ., \\ pl. Maxa
Borna 9, 50-204 Wroc{\l}aw, Poland}\\
 {\tt jerzy.lukierski@ift.uni.wroc.pl}     }
\end{center}

\vspace{0.5cm}


\vspace{0.3cm}

\end{center}

\begin{abstract}
\noindent
We present $D{=}3$ and $D{=}4$ world-line models for massive particles moving in a new type of
enlarged spacetime, with $D{-}1$ additional vector coordinates, which after quantization lead to
towers of massive higher spin (HS) free fields. Two classically
equivalent formulations are presented: one with a hybrid spacetime/bispinor variables and
a second described by a free two-twistor dynamics with constraints.
After first quantization in the $D{=}3$ and $D{=}4$ cases, the wave functions
satisfying a massive version of Vasiliev's free unfolded equations are
given as functions on the $SL(2,\mathbb{R})$ and $SL(2,\mathbb{C})$ group manifolds
respectively, which describe arbitrary on-shell momenta and spin degrees
of freedom. Further we comment on the $D{=}6$ case, and possible supersymmetric extensions
are mentioned as well. Finally, the description of interactions and the AdS/CFT
duality are briefly considered for massive HS fields.
\end{abstract}

\vspace{0.3cm}


\vspace{1cm}

-----------------------

\vspace{0.3cm}

{\footnotesize $^{\ddagger}$ On leave of absence from V.N.\,Karazin Kharkov National University, Ukraine.}

\newpage


\setcounter{footnote}{0}

\setcounter{equation}{0}
\section{Introduction}
The development of higher spin (HS) theory
was predominantly associated with massless (conformal) HS fields. One of
the important methods for the description of HS fields consists of
introducing master fields on an enlarged spacetime, which then
lead to spacetime fields with all possible values of
helicity (when the mass $m{=}0$) or spin (when $m{\neq} 0$). In particular,
a collection of $D{=}4$ massless HS fields with arbitrary
helicities was described by quantizing particles propagating
in tensorial spacetime $x_M= (x_\mu \sim x_{\alpha\dot{\beta}},
\, y_{\mu\nu} \sim (y_{\alpha\beta},\bar{y}_{\dot{\alpha}\dot{\beta}}))$
extended by commuting Weyl spinor coordinates
$y_\alpha$, $y_{\dot{\alpha}}$, $\alpha,\dot{\alpha}=1,2$
(see {\it e.g.} \cite{Frons,BL99,BLS00,V+,PST03,FI};
for the spinorial notation, see Appendix A).
It is easy to show that an equivalent particle model can be formulated
in twistor space \cite{PC72,S83}, with tensorial spacetime coordinates
eliminated by generalized Penrose incidence relations
\cite{BL99,BLS00,V+,PST03,FI}.

In this paper we consider the description of free {\it massive} HS fields,
obtained by quantization of new particle world-line model in $D=4$ generalized
spacetime $X_{\cal M}=(x_\mu,y^r_\mu)$ ($r=1,2,3$) extended by the pair of commuting Weyl spinors
$y_\alpha^i$, $\bar y_{\dot\alpha i}$ ($i=1,2$).
We recall that generalized spacetime with one auxiliary fourvector $y_\mu$
had been employed for a bilocal description of infinite massless
HS multiplets already in the seventies \cite{Fr78} (see further \cite{Sor04}). We
also add that recently, in the context of AdS/CFT duality, a similar bilocal
description has been obtained from first quantization
of a world-line biparticle model \cite{Jev11,Jev14}. In our approach, we
shall supplement spacetime with three auxiliary vectors (in $D=4$ providing 12 degrees of freedom),
but due to the phase space constraints which follow from our particle model
most of these degrees are non-dynamical.

Let us recall the considerations in \cite{BL99,BLS00,V+,PST03,FI}.
The most general $D=4$ model in $D{=}4$ tensorial spacetime describing
free HS multiplets is provided by the following action
\begin{equation}
\label{1.1}
S = \int d\tau \Big( \pi_\alpha \bar{\pi}_{\dot{\beta}} \dot{x}^{\alpha\dot{\beta}}
+a\, \pi_\alpha \pi_\beta \dot{y}^{\alpha\beta}
+ \bar{a}\,\bar{\pi}_{\dot{\alpha}} \bar{\pi}_{\dot{\beta}}
\dot{\bar{y}}^{\dot{\alpha}\dot{\beta}} +b\,\pi_\alpha \dot{y}^\alpha +
\bar{b}\,\bar{\pi}_{\dot{\alpha}} \dot{\bar{y}}^{\dot{\alpha}}\Big)\,,
\end{equation}
where $a$, $b$ are complex parameters, $\bar{\pi}_{\dot{\alpha}} \equiv
(\pi_\alpha)^*$, etc.  The model
\eqref{1.1} with $b=0$ was considered  in \cite{BLS00},
and that the last two terms ($b\neq 0$) were first introduced in
\cite{V+}. The advantage of having $b\neq 0$ is the
much simpler structure of the constraints in phase space and the
easier quantization procedure. It turns out that for $a\neq 0$
and/or $b\neq 0$ the hybrid action \eqref{1.1}
depending on tensorial spacetime and spinorial coordinates can be rewritten (modulo
boundary terms) as the one-twistor free particle model \cite{PC72,S83}
\begin{equation}
\label{1.2}
\begin{array}{rcl}
S&=& -{\textstyle\frac{1}{2}} {\displaystyle \int} d\tau \Big(\bar{Z}_A \dot{Z}^A + \textrm{h.c.}\Big) =
-{\textstyle\frac{1}{2}} {\displaystyle \int} d\tau \Big({\omega}^\alpha \dot{\pi}_\alpha -
\bar{\pi}_{\dot{\alpha}} \dot{\bar\omega}^{\dot{\alpha}}+ \textrm{h.c.}\Big)
\\ [7pt]
&=&
-{\displaystyle \int} d\tau \Big({\omega}^\alpha \dot{\pi}_\alpha
-\bar{\pi}_{\dot{\alpha}} \dot{\bar\omega}^{\dot{\alpha}}\big)+
\textrm{boundary term}
\,,
\end{array}
\end{equation}
and the $D{=}4$ twistor $Z^A$, $A=1,\dots, 4$ (conformal basic spinor) is described by
a pair of Weyl spinors
\begin{equation}
\label{1.3}
Z^A = \left(
\begin{array}{c}
\pi_\alpha \\
\bar\omega^{\dot{\alpha}} \\
\end{array}
\right)
, \qquad
\left({Z}^{ A}\right)^\dagger = \Big(
\bar{\pi}_{\dot\alpha} \,,
{\omega}^{\alpha} \Big) \,,
\end{equation}
where the conformally invariant scalar product
\begin{equation}
\label{1.4}
    \bar{Z}_A Z^A \equiv \left({Z}^{ A}\right)^\dagger g_{{A}B} Z^B = {\omega}^\alpha \pi_\alpha
    - \bar{\pi}_{\dot{\alpha}} \bar\omega^{\dot{\alpha}}
\end{equation}
is obtained by the particular choice of the anti-hermitian antisymmetric
$U(2,2)$ metric\footnote{
The choice \eqref{1.5} is used in \cite{BBCL,BCed} and has been adjusted in such a way that it
remains valid also for real $D{=}3$ twistors,
which are fundamental $Sp(4;\mathbb{R})$ spinors (see Sec.\,2.1). In $D{=}4$
this choice of the $SU(2,2)$ metric leads to purely imaginary twistor lengths (see \eqref{1.4}).
Note that the conformal groups $SO(2,\nu+2)$ ($\nu=1,2,4$) in spacetime dimensions $D=\nu+2$
are isomorphic to the $U_\alpha(4;\mathbb{K})$ groups, where $\mathbb{K}=\mathbb{R},\mathbb{C},\mathbb{H}$
are the corresponding division algebras and $U_\alpha(2n;\mathbb{K})$ are the antiunitary $\mathbb{K}$-valued matrix groups preserving
the anti-hermitian bilinear form. We have $U_\alpha(2n;\mathbb{R})\simeq Sp(2n;\mathbb{R})$,
$U_\alpha(2n;\mathbb{C})\simeq U(n,n)$ and $U_\alpha(2n;\mathbb{H})\simeq O(2n;\mathbb{H})\simeq O^\ast(4n;\mathbb{C})$
(see {\it e.g.} \cite{Tits}).}
\begin{equation}
\label{1.5}
g_{{A}B} = \left(
\begin{array}{cc}
0 & -\delta^{\dot\alpha}{}_{\dot\beta} \\
\delta_{\alpha}{}^{\beta} & 0 \\
\end{array}
\right) .
\end{equation}
The passage from the hybrid spacetime/spinor description \eqref{1.1}  to the twistorial one \eqref{1.2}
is achieved by a modified Penrose incidence relation. For the actions
\eqref{1.1} and \eqref{1.2} a suitably chosen incidence relation is:
\begin{equation}\label{1.6}
\begin{array}{rcl}
{\omega}^\alpha &=& x^{\alpha\dot{\beta}} \bar{\pi}_{\dot{\beta}} + 2a\, y^{\alpha\beta}
\pi_\beta + b\, y^\alpha  \,, \\[6pt]
\bar\omega^{\dot{\alpha}} &=& \pi_\beta x^{\beta\dot{\alpha}} +  2\bar{a}\,
{\bar y}^{\dot{\alpha}\dot{\beta}}\bar{\pi}_{\dot{\beta}}
+ \bar{b}\, \bar{y}^{\dot{\alpha} }\, .
\end{array}
\end{equation}
After inserting eqs.~\eqref{1.6} into
\eqref{1.1}, the free twistorial particle action \eqref{1.2}
follows modulo boundary terms. Besides, since $x^{\alpha\dot{\beta}}$
in the action \eqref{1.1} has to be hermitian for $x^\mu$ to be real,
inserting \eqref{1.6} in eq.~\eqref{1.4} we see that
\begin{equation}
\label{1.7}
    \bar{Z}_{A} Z^A= (2a\, \pi_\alpha \pi_\beta y^{\alpha\beta}- \textrm{h.c.})+
    (b\, \pi_\alpha y^\alpha - \textrm{h.c.})\ .
\end{equation}
Using the realization of the Poincar\'e algebra in terms of the
twistor coordinates $Z^A$, $\bar{Z}_{A}$ (see \cite{PC72,S83,Ferber}),
and using the canonical Poisson brackets (PB) following from \eqref{1.2},
it follows that in the $D=4$ massless case the helicity $h$ is given by
\begin{equation}
\label{1.8}
h = {\textstyle\frac{i}{2}}\,\bar{Z}_{A} Z^A  \ .
\end{equation}
When $a=b=0$ we obtain the Shirafuji model \cite{S83} with twistor
coordinates restricted, due to \eqref{1.7}, by the zero helicity
constraint $\bar{Z}_{A} Z^A=0$. In the twistor formulation of the Shirafuji model
\eqref{1.2}, this helicity constraint has to be added by a Lagrange multiplier.
We add that the zero value of helicity can be shifted
after quantization ($h\,\to\,\hat h$) to a non-zero one by using
various orderings for the quantized twistors in the helicity operator
$\hat h$ \cite{EisSol}.
If $a\neq 0$ and/or $b\neq 0$ the value of $h$ (see \eqref{1.7}) is not
kinematically restricted in the twistor framework and the action
describes an infinite massless multiplet with all helicities (see
{\it e.g.} \cite{BLS00}).

The HS theory formulated on generalized spacetimes with
supplementary spinorial coordinates has been employed by Vasiliev and his
collaborators for the description of interacting massless HS fields
since early 90's \cite{Vas1989,Vas1991,V+,Vas2001,GelVas2005,GelSVas2008,GelVas2010,Vas2012}.
The dynamics of free HS master fields, derived in \cite{BLS00,V+,PST03,FI}
from a first-quantization of the  particle model \eqref{1.1},
corresponds in Vasiliev theory to the simplest ``free'' choice of
general unfolded equations\footnote{The unfolding technique
consists in replacing the higher order equations
for the dynamical variables of the original system by an equivalent first order
formulation obtained by adding suitable auxiliary variables. In unfolded
HS dynamics one introduces infinitely many nondynamical auxiliary
fields; the name ``unfolding'' was introduced in \cite{Vas1994}.
It has been known for a long time that HS equations in spacetime
require higher order derivatives \cite{BBB1983,BBD1985}.}
which, in the general case, provide the description of interacting HS gauge fields
(for recent reviews se \cite{DidSk14,Vasil14}). The unfolded
equations for massive HS fields in Minkowski or $AdS$ spacetimes
have also been treated in Vasiliev framework
\cite{ProVas1997,ProVas1999,ShaVas2000,Zin2008,BIS2008,PonVas2010,BPSS2014},
although without sufficiently conclusive or general results.
In this paper we propose a new type of unfolded equations for free massive HS fields.
The novelty of our approach is the new extension of $D=3$ and $D=4$ spacetime by $D-1$
auxiliary vectors which is dual to extended momentum space
 with orthogonal Lorentz frame constraints. The mass and the spin
are introduced geometrically and, generalizing the method
for the massless HS case \cite{BL99,BLS00,V+,PST03,FI}, we
 formulate a new particle model in
two-twistor space with suitable constraints.

In this paper we describe $D{=}3$ and $D{=}4$ HS particle models
which, after first quantization, lead to free {\it massive} HS fields with
arbitrary values of spin. The application of the ideas presented in
\cite{BLS00,V+,PST03,FI} to the massive case requires the doubling
of spinor indices in the hybrid (eq.~\eqref{1.1} ) particle actions (see {\it e.g.}
\cite{FZ03,FFLM06,AIL08,MRT13}) and the enlargement of \eqref{1.2} to the free
two-twistor action (see {\it e.g.} \cite{T77,P79,H79,B84,BALM04,AFLM06,FL14}).
In our study we provide the generalizations of the actions
\eqref{1.1} and \eqref{1.2} by
incorporating the mass-shell constraints and by
introducing a suitable form of the incidence relations
linking the two-twistorial and generalized spacetime coordinates. In this way, we
obtain HS particle models with the right number of
physical phase space degrees of freedom, namely six in $D=3$ (abelian spins) and
twelve degrees of freedom in $D{=}4$ ($SU(2)$-spins). It will follow that
describing massive HS fields by an extension of the `hybrid' (eq.~{\eqref{1.1})
and  purely twistorial (eq.~\eqref{1.2}) actions produces equivalent
models with the same number of degrees of freedom.

The plan of the paper is as follows. In Sec.~\ref{D=3} we
study $D{=}3$ massive HS models. After some kinematic results about $D{=}3$
two-twistor space we describe our $D{=}3$ counterpart of the
model \eqref{1.1}. It is shown that the standard two-twistor
Shirafuji model without additional coordinates only provides
spinless massive $D{=}3$ particles (see also \cite{MRT13}). To
modify this result in order to obtain $D{=}3$ massive particles with arbitrary
spin, we introduce a spinorial action with a pair of additional
three-vector coordinates and we impose suitable mass constraints.
Further, we describe the model in phase
space and show that after solving the first class constraints
providing the unfolded equations,
we obtain a wave function on the three-dimensional $D{=}3$ spinorial Lorentz group
$SL(2;\mathbb{R})\approx\overline{SO(2,1)}$ manifold, with three
independent coordinates, two related with the three-momentum on
the mass-shell and the third with arbitrary $D{=}3$ Abelian spin values.
After introducing suitable incidence relations we obtain the two-twistor formulation
with eight-dimensional phase space restricted by one
first-class mass constraint. If we quantize such two-twistorial model
we obtain the wave function defined on the $ SL(2;\mathbb{R})$
group manifold.
After providing the realization of the $D{=}3$ spin operator
we get that the power expansion of the wave function (see \eqref{F-exp-wf})
provides in momentum space a $D{=}3$ massive infinite-dimensional
multiplet with all values of spin.

In Sec.~\ref{D=4}, the $D{=}4$ case is considered. First, we provide
variables useful in the relativistic kinematics of massive particles with spin (four-momenta,
Pauli-Luba\'nski four-vector, orthonormal bases in four-momentum
space called also Lorentz harmonics) in terms of two-twistor geometry.
Secondly, we consider the extension of the $D{=}4$ hybrid action to
two-twistor space. In the general case, the auxiliary
coordinates present in \eqref{1.1} can be enlarged by the replacements
\begin{equation}
\label{1.9}
\begin{array}{c}
x_{\alpha\dot{\beta}} \rightarrow (x_{\alpha\dot{\beta}},
y_{\alpha\dot{\beta}}^r)\,, \quad \quad y_\alpha \rightarrow
y_\alpha^i \, , \qquad r=1,2,3 \,,\qquad i,j=1,2\;,
\\[6pt]
y_{\alpha\beta} \rightarrow y_{\alpha\beta}^{ij} =
y_{\beta\alpha}^{ji}\,, \quad \quad \bar{y}_{\dot{\alpha}\dot{\beta}}
\rightarrow \bar{y}_{\dot{\alpha} \dot{\beta}}^{ij} =
\bar{y}_{ \dot{\beta}\dot{\alpha}}^{ji}\, .
\end{array}
\end{equation}
The standard Shirafuji model with spacetime coordinates $x_{\alpha\dot{\beta}}$
and a pair of spinors ($\pi_\alpha
\to \pi_\alpha^i$, $\bar{\pi}_{\dot{\alpha}} \to
\bar{\pi}_{\dot{\alpha}i}$) leads, after using the standard
incidence relation (see {\it e.g.} \cite{PC72}), to a
two-twistorial $D{=}4$ free particle model with four first class
constraints. If the two spinorial mass constraints
\begin{equation}
\label{1.10}
\mathcal{M}= \pi_\alpha^i \pi^\alpha_i +2M =0 \,,\qquad
\bar{\mathcal{M}}= \bar{\pi}_{\dot{\alpha}}^i
\bar{\pi}^{\dot{\alpha}}_i +2\bar{M}=0
\end{equation}
are further added, where $\pi^\alpha_i = \epsilon^{\alpha\beta} \epsilon_{ij}
\pi^j_\beta$ and $M$ is a complex mass parameter\footnote{
It is related with the mass parameter $m$ of the particle through
$2|M|^2=m^2$ (see also \eqref{3.6}).},
one obtains a model describing $D{=}\,4$ spinless massive particle with four first class and two second
class  constraints. To
modify the constraints that require the spin to be
zero, we introduce three additional auxiliary
four-vector coordinates $y_{\alpha\dot\beta}^r$
($r{=}1,2,3$) (see  \eqref{1.9}).  Arranging correctly the generalized incidence
relations we obtain the two-twistorial free model with one first class
and two second class constraints, which reduce the
16 twistor real coordinates $(\pi_\alpha^i,\bar{\pi}_{\dot{\alpha}i})$
(eq.~\eqref{3.1}) to $12$ physical degrees
of freedom. These new versions of the hybrid model can be
quantized and solved by using the `spinorial roots' ($\pi_\alpha^i, \bar{\pi}_{\dot{\alpha} i}$)
of the four-momenta as independent variables, which provides the reduced $D{=}4$
wave function $\psi(\pi_\alpha^i,\bar{\pi}_{\dot{\alpha}i})$.
If we take into consideration the mass constraints \eqref{1.10} we obtain that the
manifold of the spinorial coordinates is described by the group
manifold of $SL(2;\mathbb{C})$, the cover of the $D{=}\,4$ Lorentz group,
with its six real parameters being half of the twelve physical
phase space degrees of freedom that are left in the bitwistorial formulation in our model.
We show that such a wave function can be identified with the $D{=}\,4$ master field
describing an infinite-dimensional multiplet of massive HS fields with
arbitrary $D{=}\,4$ spin spectrum (for an analogy see \cite{BBTD88}).

Finally, in Sec.~\ref{beyond} we present some comments going beyond $D=3,4$,
on possible $D{=}6$ and supersymmetric extensions. Further, the
Outlook  briefly discusses how to introduce nonlinear
interactions of the massive HS multiplets and how to adapt, in our massive case,
the construction for massless HS fields in \cite{Vas1989,Vas1991,V+} which uses,
in suitably chosen dimensions,
the duality between the HS current and HS field multiplets.

The paper includes two appendices. Appendix\,A details our
conventions; Appendix\,B presents an interpretation of our $N{=}2$ $D{=}3$ spinorial
model in Sec.\,2.2 as described by an $N{=}1$ $D{=}4$ vectorial model
with the nonstandard $O(2,2)$ Lorentz group.

\section{$D=3$ bispinorial particle models and HS massive fields
from their quantization}
\label{D=3}
\setcounter{equation}{0}

\subsection{Summary of $D{=}3$ two-twistor kinematics}\label{sum}

$D{=}3$ twistors are real four-dimensional
$Sp(4;\mathbb{R})=\overline{SO(3,2)}$ spinors. We introduce a
pair of $D{=}3$ real twistors
\begin{equation}
\label{2.1}
 t^{Ai}= \left(%
\begin{array}{c}
  \lambda_\alpha^i \\
  \mu^{\alpha i} \\
\end{array}%
\right) \, , \qquad \alpha=1,2 \,,\quad i=1,2 \,,\quad A=1,\dots,4\;,
\end{equation}
with conformal-invariant scalar product\footnote{The
conformal $D=3$ twistors are null twistors.}
\begin{equation}\label{2.2}
t_A^{i}t^A_{i}=t_A^{i}\epsilon_{ij} t^{Aj}\ ,
\end{equation}
where the contravariant spinor
\begin{equation}\label{2.2a}
t_A^{i} =g_{AB}t^{Bi}
\end{equation}
is constructed using the $Sp(4;\mathbb{R})$-invariant antisymmetric metric
(see also footnote${}^{1}$)
\begin{equation}
\label{2.2b}
g_{AB} = \left(
\begin{array}{cc}
0 & -\delta^{\alpha}{}_{\beta} \\
\delta_{\alpha}{}^{\beta} & 0 \\
\end{array}
\right) .
\end{equation}

If we only employ the spinors $\lambda_\alpha^i$ we can construct the following
$D{=}3$ bilinears describing composite three-vectors in internal
$N$=2 ($i,j$=1,2) space
\begin{equation}
\label{2.3}
u_{\alpha\beta }^a = \lambda^i_\alpha (\gamma^a)_{ij}
\lambda^j_\beta \quad ,\quad a\equiv (0,r)=(0,1,2) \; ,
\end{equation}
where the $2\times 2$ matrices $(\gamma^a)_{ij}$ are internal space
$SO(2,1)$ Dirac matrices (eq.~\eqref{3gamma-3}) and form a basis
for the space of symmetric
$2{\times}2$ matrices (see Appendix A). Further, according to Penrose twistor
theory (see {\it e.g.} \cite{PC72}) we take $u_{\alpha\beta}^0
= p_{\alpha\beta}$ (three-momentum). We shall further impose the following
spinorial mass constraint
\begin{equation}
\label{2.4}
\Lambda \equiv \lambda^i_\alpha \lambda^\alpha_i +\sqrt{2}\,m =0\, , \qquad
\lambda^\alpha_i = \epsilon_{ij} \epsilon^{\alpha\beta}
\lambda^j_\beta\ ,
\end{equation}
which implies that the three-vectors \eqref{2.3} describe, after
suitable normalization $e_{\alpha\beta}^a=\frac{1}{m}\,u_{\alpha\beta}^a$,
the $D{=}3$ vectorial harmonics (see {\it  e.g.} \cite{Sok86,DKS}\footnote{The authors thank
Evgeny Ivanov for informing about the reference \cite{DKS}.}) describing
the $D{=}3$ Lorentz orthonormal vector frame
\begin{equation}
\label{2.5}
e_{\alpha\beta}^a e^{b\,\alpha\beta} = \eta^{ab} \,, \qquad \eta^{ab}=(1,-1,-1) \ .
\end{equation}
It is easy to check that the set of three-vectors $u_{\alpha\beta}^a$
has three independent degrees of freedom equal to the number of
spinorial degress of freedom constrained by the relation \eqref{2.4}.
In particular, if $a{=}b{=}0$ we obtain from \eqref{2.5} the
mass-shell condition for the $D{=}3$ momenta
\begin{equation}\label{mom-sq-3}
p_{\alpha \beta} p^{\alpha \beta}= m^2  \; ,
\end{equation}
where
\begin{equation}\label{mom-3}
p_{\alpha \beta} \equiv u^0_{\alpha \beta} =\lambda^i_{\alpha} \lambda^i_{\beta}\,.
\end{equation}

In order to describe the realizations of Lorentz group and the
Abelian scalar $D{=}3$ spin $S$ we should use all twistor
components (see \eqref{2.1}). The Lorentz algebra generators
$M_{\mu\nu}=L_{\mu\nu}+S_{\mu\nu}$, $L_{\mu\nu}=x_\mu p_\nu-x_\nu p_\mu$
are given in spinorial notation by
\begin{equation}
\label{2.6}
M_{\alpha \beta} = -{\textstyle\frac{1}{\sqrt{2}}}\, \lambda^i_{(\alpha} \mu^i_{\beta)}
\equiv -{\textstyle\frac{1}{2\sqrt{2}}}\left( \lambda^i_{\alpha} \mu^i_{\beta} +\lambda^i_{\beta} \mu^i_{\alpha}\right)
\end{equation}
and the scalar spin $S$ for the massive particle with mass $m$ is
described by the $D{=}3$ counterpart of the Pauli-Luba\'nski
operator given by ($\mu,\nu,\varrho=0,1,2$)
\begin{equation}
\label{2.7}
{\textstyle\frac12}\,\epsilon_{\mu\nu\rho} p^\mu M^{\nu\rho} = p^{\mu} M_{\mu}=
p^{\alpha\beta} M_{\alpha\beta} =mS\,,\qquad S= {\textstyle\frac12}\, \lambda_{\alpha}^{ i}
\mu^{\alpha}_{i}=  {\textstyle\frac14}\, t_A^{i}t^A_{i} \ ,
\end{equation}
where $M_\mu=\frac12\,\epsilon_{\mu\nu\lambda}M^{\nu\lambda}$ and we use the bitwistor representation of momenta \eqref{mom-3}.
We see that $D{=}3$ spin is described by the unique
nonvanishing conformal-invariant twistor norm
provided by formula \eqref{2.2}.

We shall further consider the field equations that
determine the mass and spin eigenvalues
of the $D{=}3$ Casimirs \eqref{mom-sq-3} and \eqref{2.7}.
Such field equations were also considered in quantum theory
as describing anyons, with arbitrary fractional value of $s$
(see {\it e.g.} \cite{Jack:91,SV93,CorPl}).
In the next section we obtain these equations with fixed $m$
and half-integer values of $s$ as a result of the
quantization of the new particle action.
We will not consider here the anyonic fractional spin values
that come from representations of the universal cover
$\mathbb{R}$ of the $D{=}3$ Abelian spin group ${U(1)}$.

\subsection{$D{=}3$ bispinorial generalization of the Shirafuji
model}
\label{gen}

We propose the following action for our $D{=}3$ model ($i,j=1,2$; $r=1,2$)
\begin{equation}
\label{2.8}
S^{(3)}= \int d\tau \Big[ \lambda^i_\alpha \lambda^i_\beta
\dot{x}^{\alpha \beta} + c\, \lambda^i_\alpha (\gamma^r)_{ij}
\lambda^j_\beta \dot{y}^{\alpha\beta}_r +f\, \lambda_\alpha^i
\dot{y}^\alpha_i +\ell \left(\lambda^i_\alpha \lambda_i^\alpha
+\sqrt{2}\,m \right)\Big]\ ,
\end{equation}
where $\lambda^\alpha_i = \epsilon^{\alpha \beta} \epsilon_{ij}
\lambda_\beta^j$ etc. and $\ell$ is a Lagrange
multiplier imposing the constraint $\Lambda$ in eq.~\eqref{2.4}.
The parameters $c$, $f$ may be set equal
to one by rescaling the coordinates, but we shall keep them
arbitrary in order to consider various variants of the model
(actually, the most interesting values are 0 and 1). In particular,
if we set $c{=}1$ the first two terms in \eqref{2.8} collapse into
$\lambda^i_\alpha (\gamma^a)_{ij} \lambda^j_\beta \dot{y}^{\alpha\beta}_a$
where $y^{\alpha\beta}_a=(\dot{x}^{\alpha\beta}, y^{\alpha\beta}_r$) with
$a=(0,r)=(0,1,2)$. If $c=f=0$, after using the standard
incidence relation
\begin{equation}
\label{2.9}
\mu^{\alpha i} = 2x^{\alpha\beta} \lambda_\beta^i \ ,
\end{equation}
and inserting \eqref{2.9} into \eqref{2.7}, we get $S=0$, {\it
i.e.} we obtain the model describing a spinless particle. In the
general case the incidence relation \eqref{2.9} has to
be generalized as follows\footnote{Relation \eqref{2.10} is
adjusted in order to obtain from \eqref{2.8} the free
two-twistor action (see Sec.~\ref{desc}.)}
\begin{equation}
\label{2.10}
\mu^{\alpha i} = 2x^{\alpha\beta} \lambda_\beta^i +2c\,
(\gamma^r)_{ij} y^{\alpha\beta}_r \lambda_\beta^j + f\, y^{\alpha
i}\ .
\end{equation}
After using relations \eqref{2.3} in \eqref{2.7} we obtain
\begin{equation}
\label{2.11}
S= -c\,\lambda_{\alpha}{}^i (\gamma^r)_{j}{}^i y^{\alpha\beta}_r \lambda^j_\beta
-{\textstyle\frac{1}{2}}\,f\lambda_{\alpha i} y^{\alpha i} \; ,
\end{equation}
thus, $S\neq 0$ whenever $c$ or $f$ are non-zero.

Setting $c=f=1$, the constraints defining the
momenta follow from \eqref{2.8} with the result
\begin{eqnarray}
\label{2.11a}
T_{\alpha\beta}^a = T_{\beta \alpha}^a&\equiv& p_{\alpha\beta}^a -u_{\alpha\beta}^a\approx 0 \,, \\[6pt]
G_\alpha^i &\equiv& p_{(y)}{}_\alpha^i -\lambda_\alpha^i \approx 0 \,,\label{2.11b} \\[6pt]
F^\alpha_i &\equiv& p_{(\lambda)}{}^\alpha_i \approx 0 \,.\label{2.11c}
\end{eqnarray}
Eqs.~\eqref{2.11b} and \eqref{2.11c} determine pairs of second class
constraints. After introducing for them Dirac brackets we obtain that
the variables $(y^{\alpha\beta}_a, y^\alpha_i )$, $a=0,1,2$, are canonically conjugate to
($p_{\alpha\beta}^a, \lambda^\alpha_i$) so that the non-vanishing PBs
are given by
\begin{eqnarray}
\label{2.12a}
\{ y^{\alpha\beta}_a , p_{\gamma\delta}^b \}
&=& \delta_a^b\,   \delta^{(\alpha}_\gamma \delta^{\beta)}_\delta \ ,
\\[6pt]
\{ y^\alpha_i , \lambda_\beta^j \} &=&  \delta_i^j
\delta_\beta^\alpha  \,,
\label{2.12b}
\end{eqnarray}
where we recall that
$A^{(\alpha}B^{\beta)} \equiv \frac12\left(A^{\alpha}B^{\beta}+A^{\beta}B^{\alpha} \right)$.

The model \eqref{2.8} has ten first class
constraints expressed by the formula \eqref{2.11a} and
the mass-shell constraint \eqref{2.4}. After quantization the above PB relations
can be realized in terms of $\hat y^{\alpha\beta}_a=y^{\alpha\beta}_a$,
$\hat\lambda_\alpha^i=\lambda_\alpha^i$
and the following differential operators
\begin{equation}
\label{2.13}
\hat{p}_{\alpha\beta}^a = -i
\frac{\partial}{\partial y^{\alpha\beta}_a} \ , \qquad
\hat{y}^\alpha_i = i \frac{\partial}{\partial
\lambda_\alpha^i} \ ,
\end{equation}
where, by definition, $\frac{\partial}{\partial y^{\alpha\beta}_a} \, y^{\gamma \delta}_b =
\delta_b^a\, \delta^{(\gamma}_\alpha \delta^{\delta)}_\beta$.
As a result, the quantized constraints \eqref{2.11a} after using equations
\eqref{2.3} deterimine the following three unfolded equations for the wave function
$\Phi \equiv \Phi(y^{\alpha\beta}_a ,\lambda^\alpha_i)$,
\begin{equation}
\label{2.14}
\left( i \frac{\partial}{\partial y^{\alpha\beta}_a} +
\lambda^i_\alpha (\gamma^a)_{ij}\lambda_\beta^j \right)\Phi(y^{\alpha\beta}_a ,\lambda^\alpha_i)
= 0 \,,\qquad a=(0,1,2) \;,
\end{equation}
with the following solution expressing explicitly the dependence on $y^{\alpha\beta}_{a}$,
\begin{equation}
\label{2.15}
 \Phi(y^{\alpha\beta}_a, \lambda_\alpha^i) = \exp \Big\{i
  \lambda^i_\alpha (\gamma^a)_{ij}\lambda_\beta^j\, y^{\alpha\beta}_{a}\Big\}
  \; \phi(\lambda^i_\beta) \ .
\end{equation}

Using, instead of  \eqref{2.13}, the dual differential realization in spinorial sector
\begin{equation}
\label{2.16}
\widehat{\lambda}_\alpha^i = -i\frac{\partial}{\partial
y^\alpha_i}\ ,
\end{equation}
one obtains from \eqref{2.4} a single field equation for the
reduced wave function
\begin{equation}
\label{2.17}
\left( \frac{\partial}{\partial y^\alpha_i}
\frac{\partial}{\partial y_\alpha^i} - \sqrt{2}\,m \right)
\tilde{\phi}(y^\alpha_i) =0 \ ,
\end{equation}
where
\begin{equation}
\label{2.18}
\tilde{\phi} (y^\beta_i) = \int d^4 \lambda \;
e^{i\lambda_\alpha^i y^\alpha_i} \phi(\lambda_\alpha^i) \ .
\end{equation}
In the `spinorial momentum' picture described by the spinors
$\lambda_\alpha^i$ the reduced wave function
${\phi}(\lambda_\alpha^i)$ depends on the spinorial momenta
restricted by the algebraic equation \eqref{2.4}. We see that the
wave function describing the quantum mechanical solution of the
model \eqref{2.8} depends on three degrees of freedom, two describing
the on-shell three-momenta and a third one being the (arbitrary)
value of the $D{=}3$ spin. In order to express the spin operator
\eqref{2.7} as a differential operator in spinorial momentum space
one has to consider the quantum version of the twistorial
description of model \eqref{2.8}.

Let us now compare the models \eqref{2.8} with $f{=}\,0$
and $f{\neq}\,0$ (for simplicity we set $c{=}1$). From expression \eqref{2.11}
it follows that in both models $S$ is a composite dynamical variable
that describes arbitrary $D{=}3$ spin;
however, the limit $f\rightarrow 0$
changes the structure of the constraints. Indeed, if $f{=}\,0$, those
in \eqref{2.11b} are not present; only the
constraints \eqref{2.11a}, \eqref{2.11c} and the mass-shell
constraint \eqref{2.4} appear. The alternative constraint structure is well
illustrated if the nine relations \eqref{2.11a} are
replaced by the equivalent set of nine Abelian constraints
\begin{equation}
\label{a2.23}
T^a_b \equiv  T_{\alpha\beta}^{a}u^{\alpha\beta}_{b}
=  p_{\alpha\beta}^{a} u^{\alpha\beta}_{b} - m^2
\delta^a_b\ \approx 0\ .
\end{equation}
Similarly, the four constraints \eqref{2.11c} can be replaced by four
equivalent ones as follows
\begin{equation}
\label{a2.24}
\begin{array}{rcl}
F &=& {\textstyle\frac{1}{2}}\,  \lambda_\alpha^i p_{(\lambda)}{}_{i}^\alpha
\approx 0 \ , \\ [5pt]
F_a &=& {\textstyle\frac{1}{2}}\, \lambda_\alpha^i (\gamma_a)_{i}{}^{j}
p_{(\lambda)}{}_{j}^\alpha \approx 0 \ ,
\end{array}
\end{equation}
where the $D{=}3$ gamma matrices  $(\gamma_a)_{i}{}^{j}$
satisfy the $so(1,2)$ commutation relations
\begin{equation}\label{a2.26}
[\gamma_a,\gamma_b] =- 2 \epsilon_{ab}{}^c \gamma_c \; ,
\end{equation}
with metric $\textrm{diag}(1,-1,-1)$ raising the $O(2,1)$ indices.
Using the canonical PB
$\{\lambda^i_\alpha,p_{(\lambda)}{}^\beta_j \}=\delta^\beta_\alpha \delta^i_j$,
it is seen that the thirteen new constraints ($T^b_a$, $F_a$, $F$) have
the following non-vanishing PBs:
\begin{equation}
\label{a2.27}
\begin{array}{rcl}
\{ F_a, F_b \} &=& \epsilon_{ab}{}^c F_c \ ,  \\ [5pt]
\{ F_a, T^c_b \} &=& \epsilon_{ab}{}^d T^c_d +  m^2
\epsilon_{ab}{}^c \ , \\ [5pt]
\{ F,  T^b_a \} &=&  -T^b_a - m^2 \delta^b_a\ , \\ [5pt]
\{ F, \Lambda\} &=&´ -\Lambda + \sqrt{2}  m\ .
\end{array}
\end{equation}
We see from the second and fourth equations of \eqref{a2.27} that four out
of the ten first class constraints $T^b_a$ and $\Lambda$ (eq.~\eqref{2.4})
present when $f{\neq }0$ become second class due to the
appearance of four constraints \eqref{a2.24} in the limit $f{=}0$.
These four constraints $(F_a,F)$ are second class and
describe the gauge fixing of four gauge transformations present
if $f{\neq} 0$. We can conclude that putting  $f{=}0$ in \eqref{2.8}
leads to the partial gauge fixing of four out of the ten gauge
degrees of freedom generated when $f{\neq} 0$ by the ten first class constraints
$T_{\alpha\beta}^{a}$ (or $T^a_b$) and $\Lambda$. If $f{\neq} 0$ the ten first
class constraints remove $2{\times} 10=20$ real degrees of freedom;  for
$f{=}0$ the six first class constraints plus the eight second class
remove the same number of $d.o.f.$, $2{\times}6 +8=20$. Thus, both models
have the same physical ({\it i.e.} without gauge degrees of freedom) content.
This proves the equivalence of the classical models considered
for $f{\neq} 0$ and $f{=}0$.

Finally, we point out that for $c{=}1$ our model \eqref{2.8} describes
a vectorial $SO(2,2)$-particle model, as discussed in Appendix B.

\subsection{$D{=}3$ bitwistorial description}
\label{desc}

In order to introduce the twistor coordinates \eqref{2.1}, we
insert in \eqref{2.8} the generalized incidence relation \eqref{2.10}.
Modulo boundary terms, we obtain for $c{\neq} 0$ and/or $f{\neq} 0$ the following
twistorial free action with $Sp(4,\mathbb{R})$ $D{=}3$ twistorial metric \eqref{1.5} is obtained:
\begin{equation}
\label{2.19}
\tilde{S}^{(3)} = \int d\tau \Big[ \lambda^i_\alpha\dot{\mu}^{\alpha i}  + \ell
\left(\lambda^i_\alpha \lambda_i^\alpha +\sqrt{2}\,m\right)\Big]\ .
\end{equation}
The action \eqref{2.19} describes an infinite tower of $D{=}3$
free massive particles with any spin (see {\it e.g.} \cite{MRT13}).
Let us prove it.

The action \eqref{2.19} describes a system with canonical
variables $\mu^{\alpha i}$ and $\lambda^i_\alpha$,
$\{\mu^{\alpha i}, \lambda^j_\beta \}=\delta^{ij}
\delta^\alpha_\beta$, and the constraint  \eqref{2.4}
which generates the gauge transformations in bitwistor space.
Let us fix this gauge freedom by the constraint
\begin{equation}
\label{constr-fix}
G=\lambda_\alpha^i\mu^{\alpha i} \approx 0\,,\qquad \{\Lambda, G \}=2\sqrt{2}\,m-2\Lambda\,.
\end{equation}
Introducing Dirac brackets incorporating the constraints
$\Lambda \approx 0$ and $G \approx 0$ we obtain that they become
strong and we get the following Dirac brackets for the twistor variables
\begin{equation}
\label{DB-twi}
\begin{array}{rcl}
\{\lambda^i_\alpha, \lambda^j_\beta \}_{\ast}&=&0\,,\\ [5pt]
\{\mu^{\alpha i}, \lambda^j_\beta \}_{\ast}&=&\delta^{ij}
\delta^\alpha_\beta+\textstyle{\frac{1}{\sqrt{2}m}}\,\lambda_i^\alpha \lambda^j_\beta\,,\\ [5pt]
\{\mu^{\alpha i}, \mu^{\beta j} \}_{\ast}&=&-\textstyle{\frac{1}{\sqrt{2}m}}\,(\lambda_i^\alpha \mu^{\beta j}
-\lambda_j^\beta \mu^{\alpha i})\,.
\end{array}
\end{equation}

A quantum realization of the algebra  \eqref{DB-twi}
with $\hat\lambda\hat\mu$ ordering is the following
\begin{equation}
\label{2.21}
\hat\lambda^i_\alpha =\lambda^i_\alpha \quad ,\quad \hat\mu^{\alpha i} =
i\,\frac{\partial}{\partial\lambda_\alpha^i}
+\frac{i}{\sqrt{2}m}\lambda_i^\alpha\,\lambda^j_\beta\frac{\partial}{\partial\lambda^j_\beta} \,.
\end{equation}
We point out that the second class constraints \eqref{2.4} and \eqref{constr-fix}
are fulfilled in the strong sense,
{\it i.e.} $\hat G=\hat\lambda_\alpha^i\hat\mu^{\alpha i} \equiv 0$.
If we use the formulae \eqref{2.21}, the spin operator \eqref{2.7} is realized as follows
\begin{equation}
\label{2.23}
\hat{S}=  {\textstyle\frac12}\, \hat\lambda_{\alpha}^{i}\hat\mu^{\alpha}_{i}=
 {\textstyle\frac i2}\, \epsilon_{ij}\, \lambda_{\alpha}^i\, \frac{\partial}{\partial \lambda_{\alpha}^{j}} \ .
\end{equation}
Our aim will be to decompose the Fourier transform \eqref{2.18}
of the reduced wave function $\tilde{\phi}(y^\alpha_i)$ satisfying
eq.~\eqref{2.17} into a superposition of momentum-dependent eigenfunctions of
the operator \eqref{2.23} (see eqs.~\eqref{mom-su}, \eqref{F-exp-wf-cov-gen} below).

Due to the mass constraint \eqref{2.4}, the real $2{\times}2$ matrices $h$ with elements
\begin{equation}
\label{sl-2r}
h_\alpha{}^i=2^{1/4}m^{-1/2} \lambda_\alpha{}^i
\end{equation}
have determinant equal to one, characterize the $SL(2;\mathbb{R})$ group manifold
and  describe real spinorial $D{=}3$
harmonics \cite{DKS} (note the algebra isomorphisms
$sl(2;\mathbb{R})\sim su(1,1)\sim sp(2;\mathbb{R})$).
The corresponding $SU(1,1)$ matrix is obtained by the complex
similarity transformation
\begin{equation}
\label{sim-tr}
g=U\,h\,U^{-1}\,,\qquad U=e^{-i\pi\sigma_1/4}\,,
\end{equation}
with matrix elements
\begin{equation}
\label{su-11}
g=
\left(
\begin{array}{cc}
a & \bar b \\
b & \bar a \\
\end{array}
\right),\qquad
|a|^2-|b|^2=1\;,\quad g\in SU(1,1) \; ,
\end{equation}
where
\begin{equation}
\label{ab-lambda}
a={\textstyle\frac12}\left[h_1{}^1+h_2{}^2+i(h_1{}^2-h_2{}^1) \right]
,\qquad
b={\textstyle\frac12}\left[h_1{}^2+h_2{}^1-i(h_1{}^1-h_2{}^2) \right]\,.
\end{equation}
In terms of the variables \eqref{ab-lambda} the spin
operator \eqref{2.23} takes the form
\begin{equation}
\label{hat-S}
\hat{S}= -\frac12\left(a\,\frac{\partial}{\partial a}+
b\,\frac{\partial}{\partial b}- \bar a\,\frac{\partial}{\partial \bar a}-
\bar b\,\frac{\partial}{\partial \bar b} \right) .
\end{equation}
The matrix $g_\alpha{}^i$ in \eqref{su-11} describes $SU(1,1)$ spinorial harmonics,
where first column $g_\alpha{}^1=\left(\!\!
\begin{array}{c}
a \\
b
\end{array}
\!\!
\right)
$
(second
$g_\alpha{}^2=\left(\!\!
\begin{array}{c}
\bar b \\
\bar a \\
\end{array}
\!\!\right)
$)
describes a $SU(1,1)$ spinor with spin eigenvalue $s=-\frac12$ ($s=\frac12$).

One can introduce the natural parametrization of the $SU(1,1)$
matrices \eqref{su-11} \cite{Vil68}
\begin{equation}
\label{su11-par}
a= \cosh (r/2)e^{i(\psi+\varphi)/2}\,,\qquad
b= \sinh (r/2)e^{i(\psi-\varphi)/2}\,,
\end{equation}
where
\begin{equation}
\label{su11-par-int}
0\leq \varphi\leq 2\pi\,,\qquad 0<r<\infty\, ,\qquad
-2\pi\leq \psi< 2\pi\, .
\end{equation}
In terms of the angle $\psi$, the operator \eqref{hat-S} takes the simple form
\begin{equation}\label{hat-S1}
\hat{S}=  i\,\frac{\partial}{\partial \psi}\,
\end{equation}
{\it i.e.}, it describes the $D{=}3$ $U(1)$ spin.

After the transformation \eqref{sl-2r}, the twistorial wave function $\Psi(g)$
is defined on $SU(1,1)$. The $SU(1,1)$ regular representation is given
by its action of on the (wave) functions $\Psi(g)$ defined on the
$SU(1,1)$ manifold. To obtain the Hilbert space of the quantized model \eqref{2.19} we may use
the theory of special functions on matrix group manifolds (see {\it e.g.}~\cite{Vil68})
and require that the wave function $\Psi(g)=\Psi(\varphi,r,\psi)$
is square-integrable, $\int|\Psi(g)|^2 dg<+\infty$, $dg=\sinh r\,dr \,d\varphi\, d\psi$.
Due to eq.~\eqref{su11-par},
the wave function satisfies the periodicity conditions
\begin{equation}
\label{prop-wf}
\Psi(\varphi,r,\psi)= \Psi(\varphi+4\pi,r,\psi)= \Psi(\varphi,r,\psi+4\pi)=
\Psi(\varphi+2\pi,r,\psi+2\pi)\,,
\end{equation}
which eliminate the anyonic quantum states with arbitrary
fractional spin.

   One can use the double Fourier expansion
\begin{equation}
\label{F-exp-wf}
\Psi(\varphi,r,\psi)= \sum_{k,n=-\infty}^{\infty} f_{kn}(r)\,e^{-i(k\varphi+n\psi)}=
\sum_{n=-\infty}^{\infty}e^{-in\psi}F_{n}(r,\varphi)\,,
\end{equation}
where
$
F_{n}(r,\varphi)\equiv \sum\limits_{k=-\infty}^{\infty} f_{kn}(r)\,e^{ik\varphi}
$
($n$ is fixed). The summation is over all pairs $(k,n)$ such that the numbers $k$
and $n$ are both integer or half-integer.
The eigenvalues of the operator $\hat{S}$ defined by \eqref{hat-S1} coincide with parameter $n$
in the expansion \eqref{F-exp-wf}. As a result, the spin in our model
takes {\it quantized} integer and half-integer values.
The functions $F_{n}(r,\varphi)$ describe states with definite $D{=}3$ spin equal to $n$.
The $r$-dependent fields in \eqref{F-exp-wf} are expressed by
\begin{equation}
\label{inv-exp-wf}
f_{kn}(r)= \frac{1}{8\pi^2}\int\limits_{-2\pi}^{2\pi}\int\limits_{0}^{2\pi}
d\varphi\, d\psi\, e^{i(k\varphi+n\psi)}\,\Psi(\varphi,r,\psi)
\end{equation}
and the Plancherel formula gives
\begin{equation}
\label{Pl-form}
\frac{1}{8\pi^2}\int\limits_{-2\pi}^{2\pi}\int\limits_{0}^{2\pi}\int\limits_{0}^{\infty}
d\varphi\, d\psi\,dr\,|\Psi(\varphi,r,\psi)|^2\, \sinh r=
\sum_{k,n=-\infty}^{\infty}\int\limits_{0}^{\infty}dr\, |f_{kn}(r)|^2\, \sinh r \,.
\end{equation}
Square integrable functions $f_{kn}(r)$ have an (integral) expansion on the matrix
elements of the $SU(1,1)$ infinite-dimensional unitary representations
(see \cite{Vil68,Bar47} for details).
Using \eqref{mom-3}  and  \eqref{sl-2r}, \eqref{su11-par} we obtain that
\begin{equation}
\label{mom-comp}
\begin{array}{rcl}
p_0 &=& m\left(a\bar a+ b\bar b\right)=m\,\cosh r \,,\\ [6pt]
p_1 &=& im\left(a\bar b-b\bar a \right)=-m\,\sinh r\,\sin \varphi\,,\\ [6pt]
p_2 &=& m\left(a\bar b+b\bar a \right)=m\,\sinh r\,\cos \varphi\, ;
\end{array}
\end{equation}
where $p_0^2-p_1^2-p_2^2 = m^2$.
We see that the on-shell momentum components \eqref{mom-comp} do not
depend on the angle $\psi$ and thus define the coset manifold
$SU(1,1)/U(1)$, the hyperboloid which is the base manifold of the
(trivial) $U(1)$-fibration of $SU(1,1)$.
 The wave function \eqref{prop-wf}
with the Fourier expansion \eqref{F-exp-wf}
in the $U(1)$ $\psi$-variable
describes an infinite-dimensional tower of $D$=3 higher spin fields.

The coefficient fields in the expansion in \eqref{F-exp-wf}
are defined on the coset $SU(1,1)/U(1)$
as functions of the on-shell three-momenta $p_\mu$,
\begin{equation}
\label{F-fix-n1}
F_{n}(r,\varphi)=\tilde F_{n}(p_\mu;m)
\end{equation}
and
\begin{equation}
\label{F-fix-n1a}
f_{kn}(r)=\tilde f_{kn}(p_0;m)\,.
\end{equation}
Let us analyze the expansion \eqref{F-exp-wf} in a Lorentz covariant form.

We recall that the transformation \eqref{sim-tr} describes
the isomorphism between $SL(2;\mathbb{R})$ and $SU(1,1)$ matrix group (see, for example, \cite{Ruhl}).
Using eq.~\eqref{sim-tr} one can transform $D{=}3$ spinors and  $\gamma$-matrices from
Majorana (real) representation to a complex representation.
We get in such a way the $D{=}3$ framework which uses the
$SU(1,1)$ spinor coordinates \footnote{
We use the index $\alpha=1,2$ for the real $SL(2;\mathbb{R})$ as well as for
the complex $SU(1,1)$ spinors since it is a Dirac spinor index in different
realizations of the $D{=}3$ $\gamma$-matrices.
Note that the reality of a $SL(2;\mathbb{R})$ spinor $\chi=\bar{\chi}$ implies the validity
of $D{=}3$ $SU(1,1)$ Majorana condition $\psi^\dagger\gamma_0=\psi^T C$ for the $SU(1,1)$ spinor
$\psi=U\chi$, where in accordance with \eqref{3gamma-c} $\gamma_0=i\sigma_3$, $C=i\sigma_2$.
}
\begin{equation}
\label{sp-su11}
\xi_\alpha=
\sqrt{m}\left(
\begin{array}{c}
a \\
b \\
\end{array}
\right) ,\qquad
{\bar\xi}^\alpha=(\xi_\alpha)^\dagger=\sqrt{m}\left(\bar a, \bar b\right)\,,\qquad
\bar\xi^\alpha(\sigma_3)_\alpha{}^\beta \xi_\beta=m\,.
\end{equation}
In the variables \eqref{sp-su11}
the $D{=}3$ spin operator \eqref{hat-S} takes the form
\begin{equation}
\label{hat-S1a}
\hat{S}= {\textstyle\frac12}\,
\left({\bar\xi}^\alpha\,\frac{\partial}{\partial {\bar\xi}^\alpha}-
\xi_\alpha\,\frac{\partial}{\partial \xi_\alpha}
\right) .
\end{equation}
We find easily that in terms of the $SU(1,1)$ spinors \eqref{sp-su11} the
three-momentum \eqref{mom-comp} is given by
\begin{equation}
\label{mom-su}
p_\mu=\widetilde{\xi}^\alpha(\gamma_\mu)_{\alpha\beta} \xi^\beta\,,
\end{equation}
where $\widetilde{\xi}^\alpha=\bar\xi^\beta (\gamma_0)_\beta{}^\alpha$ is the
Dirac conjugated spinor, $\xi^\beta=\epsilon^{\beta\alpha}\xi_\alpha$,
$(\gamma_\mu)_\alpha{}^\beta$ are Dirac $\gamma$-matrices in the
complex $SU(1,1)$ representation \eqref{3gamma-c},
$(\gamma_\mu)_{\alpha\beta}=\epsilon_{\beta\gamma}(\gamma_\mu)_{\alpha}{}^{\gamma}$
and  $p_\mu=\widetilde{\xi}^\alpha(\gamma_\mu)_{\alpha\beta} \xi^\beta=
-\widetilde{\xi}^\alpha (\gamma_\mu)_{\alpha}{}^\beta\xi_\beta \equiv -\tilde\xi \gamma_\mu \xi$.
Eq.~\eqref{mom-su} is the $D{=}3$ counterpart of the standard Penrose formula for
the four-momenta in the $D$=4 case, in which the $D{=}4$ $SL(2;\mathbb{C})$ Weyl spinors
have been replaced by $D{=}3$ $SU(1,1)$ spinors.

Using relations \eqref{su11-par} and \eqref{sp-su11} we can write down
the expansion \eqref{F-exp-wf}
in the covariant form
\begin{equation}
\label{F-exp-wf-cov-gen}
\Psi(\xi,\bar\xi)= \sum_{N,K=0}^{\infty} {\xi}_{\alpha_1}\ldots {\xi}_{\alpha_K}\,
\bar{\xi}^{\beta_1}\ldots \bar{\xi}^{\beta_N}\, \psi_{\beta_1\ldots\beta_N}^{\alpha_1\ldots\alpha_K}(p_\mu)\,,
\end{equation}
where $\psi_{\beta_1\ldots\beta_N}^{\alpha_1\ldots\alpha_K}(p_\mu)$
is the covariant counterpart of the functions $\tilde F_{n}(p_\mu;m)$
where $\frac{N-K}{2}=n$ (see eqs.~\eqref{F-exp-wf-cov-gen}, \eqref{sp-su11},
\eqref{su11-par} and \eqref{F-exp-wf}).

We note that the $SU(1,1)$ spinorial formalism is more convenient for
the description of spin states than the $SL(2;\mathbb{R})$ framework
because it diagonalizes the spin eigenvalues.
Formally the wave function \eqref{F-exp-wf-cov-gen} (or the reduced
wave function $\phi(\lambda_\alpha^i)$ in \eqref{2.18}), after using
\eqref{2.4}, can be written as follows
\begin{equation}
\label{F-exp-wf-cov}
\phi(\lambda)= \sum_{N=0}^{\infty}
\lambda_{\alpha_1}^{i_1}\ldots \lambda_{\alpha_N}^{i_N}\,
\tilde\psi^{\alpha_1\ldots\alpha_N}_{i_1\ldots i_N} (p_\mu)\, .
\end{equation}
However, the monomials $\lambda_{\alpha_1}^{i_1}\ldots \lambda_{\alpha_N}^{i_N}$
are not eigenvectors of the spin operator \eqref{2.23}.

We point out that the expansions \eqref{F-exp-wf-cov-gen} include
both states with positive ($n>0$) and negative ($n<0$) spin values
and that it is infinitely degenerate because a spin $n$ is generated by all
monomials ${\xi}_{\alpha_1}\ldots {\xi}_{\alpha_K}\,
\bar{\xi}^{\beta_1}\ldots \bar{\xi}^{\beta_N}$ such that $n=\frac{N-K}{2}$.
One can remove the degeneracy in $N,K$ for a given $n$ by
projecting on the spaces with definite sign of spin if we consider
anti-holomorphic wave functions satisfying the condition
\begin{equation}
\label{ahol}
\frac{\partial}{\partial\xi_{\alpha}}\,\Psi(\xi,\bar\xi)=0 \; .
\end{equation}
A solution of \eqref{ahol} is provided by the power serie
\begin{equation}
\label{F-exp-wf-cov-+}
\Psi^{(+)}(\bar\xi)=
\sum_{N=0}^{\infty} \bar{\xi}^{\alpha_1}\ldots \bar{\xi}^{\alpha_N}\,
\psi^{(+)}{}_{\alpha_1\ldots\alpha_N}(p_\mu) \; ,
\end{equation}
which depends only on ${\bar \xi}$ and contains only positive spins.

Alternatively, we may impose the condition
\begin{equation}
\label{hol}
\frac{\partial}{\partial\bar\xi^{\alpha}}\,\Psi(\xi,\bar\xi)=0 \ ,
\end{equation}
which can also be interpreted as another $SU(1,1)$ harmonic
expansion condition.

The spacetime dependent fields are obtained in the standard way by means
of a  generalized Fourier transform with exponent
$e^{i p_{\mu} x^{\mu}}=e^{-i (\tilde\xi\gamma_\mu\xi) \,x^{\mu}}$
and measure $\mu^3(\xi)=d^4\xi\, \delta(\bar\xi\sigma_3\xi -m)$
(see eq.~\eqref{sp-su11}). We get in such a way the Fourier-twistor transform
for $D{=}3$ massive fields. The corresponding spacetime fields are then given by
\begin{equation}
\label{comp+}
\phi^{(+)}_{\alpha_1\ldots\alpha_N}(x) =
\int \mu^3(\xi)\, e^{-i (\tilde\xi\gamma_\mu\xi) \,x^{\mu}}
\xi_{\alpha_1}\ldots \xi_{\alpha_N}\,\Psi^{(+)}(\xi)\ .
\end{equation}
The fields \eqref{comp+} are symmetric with respect to their
spinorial indices and satisfy the $D{=}3$ Bargmann-Wigner equations
\begin{equation}
\label{D-comp+}
\partial_\mu(\gamma^\mu)_{\beta}{}^{\alpha_1}\phi^{(+)}_{\alpha_1\alpha_2\ldots\alpha_N}-
m \phi^{(+)}_{\beta\alpha_2\ldots\alpha_N}=0\; ,
\end{equation}
where the $\gamma$-matrices are taken in the complex $SU(1,1)$
representation \eqref{3gamma-c}.

The negative ($n< 0$) spin (helicity)
states  are described by
the holomorphic twistor wave function
\begin{equation}
\label{F-exp-wf-cov--}
\Psi^{(-)}(\xi)= \sum_{N=0}^{\infty}
\xi_{\alpha_1}\ldots \xi_{\alpha_N}\, \psi^{(-)}{}^{\alpha_1\ldots\alpha_N}(p_\mu)\,,
\end{equation}
which is a solution of equation \eqref{hol}.
The twistor transform can be obtained by the complex conjugation of \eqref{comp+}
\begin{equation}
\label{comp-}
\phi^{(-)}{}^{\alpha_1\ldots\alpha_N}(x) = \int \mu^3(\xi)\, e^{i (\tilde\xi\gamma_\mu\xi) \,x^{\mu}}
\bar\xi^{\alpha_1}\ldots \bar\xi^{\alpha_N}\,\Psi^{(-)}(\xi)
\end{equation}
and defines spacetime fields with symmetric spinorial indices that satisfy
the Bargmann-Wigner equations \eqref{D-comp+} with $m\,\to\,-m$.

\section{$D{=}\,4$ bispinorial models and HS massive fields}
\label{D=4}
\setcounter{equation}{0}

\subsection{Summary of $D{=}\,4$ two-twistor kinematics}
\label{4summ}

The standard $D{=}\,4$ Penrose twistors are complex
four-dimensional $SU(2,2)= \overline{SO(4,2)}$ spinors
$Z^{Ai}$, $\bar Z_{Ai}$ that can be expressed by two pairs of two-component Weyl
spinors ($\pi_\alpha^i,\bar\omega^{\dot{\alpha}i}$)
\begin{equation}
\label{3.1}
Z^{Ai} = \left(%
\begin{array}{c}
\pi^i_\alpha \\
\bar{\omega}^{\dot{\alpha} i} \\
\end{array}%
\right) , \qquad
\left(Z^{Ai}\right)^\ast \equiv \left(%
\begin{array}{c}
\bar{\pi}_{\dot{\alpha} i} \\
\omega^{\alpha}_{ i} \\
\end{array}%
\right),\qquad \bar Z_{Ai}=
\left( \omega^{\alpha}_{ i}, -\bar{\pi}_{\dot{\alpha} i}\right)
\end{equation}
where $\bar{\pi}_{\dot{\alpha} i}=({\pi^i_\alpha})^\ast$, $\omega^{\alpha}_{ i}=
({\bar\omega^{\dot\alpha i}})^\ast$.
One can introduce four conformal-invariant scalar products ($a=0,1,2,3$)
\begin{equation}
\label{3.2}
S_{i}{}^{j}= \bar{Z}_{Ai} Z^{Aj}\qquad \textrm{or} \qquad S^a =
Z^{Ai}(\sigma^a)_{i}{}^{j} \bar{Z}_{Aj} \ ,
\end{equation}
where the hermitian $2{\times}2$ matrices above $\sigma^a$ are defined
in Appendix A and act in the internal bidimensional space.

Using the complex Weyl spinors $\pi_\alpha^i,
\bar{\pi}_{\dot{\alpha}i}$ we can define the following set of real
composite four-vectors
\begin{equation}
\label{3.3}
u_{\alpha\dot\beta}^a = \pi_\alpha^i (\sigma^a)_{i}{}^{j}
\bar{\pi}_{\dot{\beta}j} \quad ,\quad a=0,1,2,3 \quad ,
\end{equation}
which for $a{=}0$ give the Penrose formula for the composite
four-momentum \cite{PC72}
\begin{equation}
\label{3.4}
u_{\alpha\dot\beta}^0\equiv p_{\alpha\dot{\beta}} =
\pi_\alpha^i \bar{\pi}_{\dot{\beta}i}\equiv
{\textstyle\frac{1}{\sqrt{2}}} \,\sigma^\mu_{\alpha\dot\beta}p_\mu\ .
\end{equation}
We shall impose (see \eqref{1.10}) two complex spinorial mass constraints
by means of the complex mass parameter $M=M_1+iM_2$. From
\eqref{3.4} and \eqref{1.10} it follows easily that
\begin{equation}
\label{3.5}
p_{\alpha\dot{\beta}} p^{\alpha\dot{\beta}} = p_\mu p^\mu = 2|M|^2\,,
\end{equation}
{\it i.e.}
\begin{equation}
\label{3.6}
|M|^2 = {\textstyle\frac{1}{2}} \,m^2\ ,
\end{equation}
where $m$ is the mass of the particle. Using further the real
four-vector notation
\begin{equation}
\label{xxx}
u_{\mu}^{a} = {\textstyle\frac{1}{\sqrt{2}} }\,(\sigma_\mu)^{\alpha
\dot{\beta}} u_{\alpha \dot{\beta}}^{a} \quad,
\quad  e^a_\mu= {\textstyle\frac{1}{m}}\, u_{\mu}^{a} \, ,
\end{equation}
it follows that ({\it cf.}~\eqref{2.5})
\begin{equation}
\label{3.7}
 u_{\mu\,a} u^{\mu}{}_b = m^2 \eta_{ab} \quad , \quad e_{\mu\,a} e^{\mu}{}_b = \eta_{ab}
 \quad , \quad  \eta_{ab}=(1,-1,-1,-1) \ .
 \end{equation}
The four-vectors $e^a_\mu$ in eqs.~\eqref{xxx}, \eqref{3.7}
describe an orthonormal vectorial Lorentz frame defining $D{=}\,4$ vectorial Lorentz
harmonics; the spinors $\sqrt{\frac{2}{m}}\,\pi_\alpha^i,
\sqrt{\frac{2}{m}}\,\bar{\pi}_{\dot{\alpha}i}$ constitute
a pair of complex-conjugated spinorial $D{=}\,4$ Lorentz harmonics
\cite{Band,DGS92,FedZim}.

The two-twistorial realization of the $D{=}\,4$ Poincar\'e algebra
$P_\mu\simeq P_{\alpha\dot{\beta}}$, $M_{\mu\nu} \simeq
(M_{\alpha\beta}, M_{\dot{\alpha}\dot{\beta}})$ can be expressed
in terms of the twistor components (eq.~\eqref{3.1}) as
follows\footnote{In \eqref{3.8} we assume the canonical
quantization rules for the twistor variables; see also Sec.\,3.3.}
\cite{PC72}.
\begin{equation}
\label{3.8}
P_{\alpha\dot{\beta}} = \pi_\alpha^i
\bar{\pi}_{\dot{\beta}i}\ ,\qquad
M_{\alpha\beta} =
\pi_{(\alpha}^i \omega_{\beta)i}  \ ,\qquad
M_{\dot{\alpha}\dot{\beta}} =
\bar{\omega}_{(\dot{\alpha}}^i \bar{\pi}_{\dot{\beta})i} \ .
\end{equation}
The Pauli-Luba\'nski four-vector  $W_\mu$ describing the $D=4$
relativistic spin,
\begin{equation}\label{3.9}
W_\mu = {\textstyle\frac{1}{2}} \,\epsilon_{\mu\nu\rho\sigma} P^\nu
M^{\rho\sigma}\ ,
\end{equation}
can be written after using expressions \eqref{3.8} and \eqref{3.2} as
an expression in twistorial coordinates as follows
\begin{equation}
\label{3.10}
W^{\alpha\dot{\beta}} = S_r\, u^{\alpha\dot{\beta}}_{r} \,,\qquad r=1,2,3\,,
\end{equation}
where
\begin{equation}\label{Sr-s}
S_r=-{\textstyle\frac{i}{2}}\left(\pi^i_{\alpha}\omega^{\alpha}_{j}-
\bar\pi_{\dot\alpha j}\bar\omega^{\dot\alpha i}\right)(\sigma_r)_i{}^j\,,\qquad r=1,2,3\,.
\end{equation}
Further, using the relations
\eqref{1.10}, \eqref{3.5} and \eqref{3.6} it follows that
\begin{equation}
\label{3.12}
W^\mu W_\mu =-m^2 {\vec{S}}^{\,2} \ ,\qquad {\vec{S}}^{\,2} \equiv S_r S_r \ .
\end{equation}
After quantization, as it is shown in Sec.\,3, we obtain the well known relativistic
spin square spectrum with ${\vec{S}}^{\,2}$ replaced by $s(s+1)$
($s=0,\frac{1}{2},1,\dots$). We observe that the covariant
generators $S_r$, which (see \eqref{3.10} and \eqref{3.7}) can be
expressed as
\begin{equation}
\label{3.13}
S_r = - \frac{1}{m^2}\, u^{\alpha\dot{\beta}}_{r}
W_{\alpha\dot{\beta}} \
\end{equation}
and describe the  $su(2)$ spin algebra in a Lorentz frame-independent
way.

\subsection{D{=}\,4 bispinorial generalization of Shirafuji model}
\label{4gen}

Following the choice made in the $D{=}3$ case (see \eqref{2.8}), we
shall generalize the standard $D{=}\,4$ bispinor Shirafuji action by
adding three additional terms  depending on the supplementary
four-vectors ${y}^{\mu}_{r}$ ($r=1,2,3$) and on the spinorial kinetic
terms, plus the pair of spinorial mass shell constraints
$\mathcal{M}, \bar{\mathcal{M}}$ in eq.~\eqref{1.10}:
\begin{eqnarray}
\label{3.14}
S^{(4)} &=& \int d\tau \Big[ \pi_\alpha^i
\bar{\pi}_{\dot{\beta}i}\, \dot{x}^{\alpha\dot{\beta}} + c\,
\pi_\alpha^i (\sigma^r)_{i}{}^{j} \bar{\pi}_{\dot{\beta}j}\,
\dot{y}^{\alpha\dot{\beta}}_{r} + f\, \pi_\alpha^i
\dot{y}^\alpha_i + \bar f\, \bar\pi_{\dot\alpha i}
\dot{\bar y}^{\alpha i} \nonumber\\
& &  \qquad + \rho\left(\pi_\alpha^i \pi^\alpha_i+2M\right) +
\bar\rho\left(\bar\pi_{\dot\alpha}^i \bar\pi^{\dot\alpha}_i+2\bar M\right) \Big]\,.
\end{eqnarray}
In \eqref{3.14} we have extended spacetime
$x^{\mu} \,{=}\, \frac{1}{\sqrt{2}} \,(\sigma^\mu)_{\alpha\dot{\beta}}
x^{\alpha\dot{\beta}}$ by the three supplementary real four-vectors
$y^{\mu}_{r} \,{=}\, \frac{1}{\sqrt{2}} \,(\sigma^\mu)_{\alpha\dot{\beta}}
y^{\alpha\dot{\beta}}_{r}$. The parameter $c$ is real and $f$ is
complex; $\rho$ and $\bar\rho$ are complex Lagrange multipliers
that impose the spinorial mass shell constraints.

  When $c=f=0$, $S^{(4)}$ describes the standard bispinorial Shirafuji
model, with the pair of standard incidence relations
\begin{equation}
\label{3.15}
\bar{\omega}^{\dot{\alpha} i}
= \pi^i_\beta x^{\beta\dot{\alpha}}\ , \qquad \omega^{\alpha}_{i} =
x^{\alpha\dot{\beta}} \bar{\pi}_{\dot{\beta}i} \ .
\end{equation}
The reality of the spacetime coordinates $x^\mu$ implies, after
multiplying the first equation above on the right side by $A_i{}^{j}
\bar{\pi}_{\dot{\alpha}j}$ and the second one on
the left side by $\pi_\alpha^j A^{ji}$, the constraint
\begin{equation}
\label{3.16}
\pi_\alpha^j A_j{}^{i} \omega^{\alpha}_{i} -
\bar{\omega}^{\dot{\alpha}i} A_{i}{}^{j} \bar{\pi}_{\dot{\alpha}j} = 0 \ ,
\end{equation}
which depends on the arbitrary hermitian $2\times 2$ matrix $A_i{}^j$, {\it i.e.}
$(A_i{}^j)^\dagger= A_j{}^i$. Using the  $\sigma^a$ basis of
$2{\times} 2$ hermitian matrices (Appendix A), eq.~\eqref{3.16} gives
the following four linearly independent constraints
($a=(0;r)=(0;1,2,3)$)
\begin{equation}
\label{3.17}
S_a \equiv -{\textstyle\frac{i}{2}}\left[\pi_\alpha^j (\sigma_a)_j{}^{i} \omega^{\alpha}_{i} -
\bar{\omega}^{\dot{\alpha}i} (\sigma_a)_{i}{}^{j} \bar{\pi}_{\dot{\alpha}j}\right] = 0\ ,
\end{equation}
which can also be expressed by the four conformal scalar products of the twistors
$Z_A^i,\bar{Z}_A^i$,
\begin{equation}
\label{3.18}
S_a \equiv -{\textstyle\frac{i}{2}}\,Z^{Ai} (\sigma_a)_{i}{}^{j} \bar{Z}_{Aj} = 0 \,  .
\end{equation}
If relation \eqref{3.18} is valid, we see that the twistors
generated by the incidence relation \eqref{3.15} are null
twistors located on the null plane.
The four constraints  \eqref{3.18} and two spinorial mass constraints \eqref{1.10}
provide four first class constraints and two of second class (see also \cite{MRT13}),
{\it i.e.} if $c=f= 0$ we obtain $16-2{\times}4-2=6$ physical degrees of freedom
describing the physical phase space of massive spinless particle.

In the general case when $c{\neq}\, 0$ and $f{\neq}\, 0$ the proper
generalization of the incidence relations is the following
\begin{equation}
\label{3.19}
\begin{array}{rcl}
\bar{\omega}^{\dot{\alpha} i} &=& \pi^i_\beta x^{\beta\dot{\alpha}}
+ {c}\, \pi^j_\beta (\sigma^r)_{j}{}^{i}
y^{\beta\dot{\alpha}}_{r}+ \bar{f}\, \bar{y}^{\dot{\alpha} i}\; , \\[6pt]
\omega^{\alpha}_{i} &=& x^{\alpha\dot{\beta}}
\bar{\pi}_{\dot{\beta}i} + c\, y^{\alpha\dot{\beta}}_r
(\sigma^r)_{i}{}^{j} \bar{\pi}_{\dot{\beta}j} + fy^{\alpha}_{i} \ .
\end{array}
\end{equation}
Repeating the derivation of the constraints \eqref{3.16}, we
obtain in place of the formulae \eqref{3.17} the
following relations ($i,j=1,2\,;\,r=1,2,3$):
\begin{equation}
\label{3.20}
\begin{array}{rcl}
S_0 &=& -{\textstyle\frac{i}{2}}\left(f\, \pi_{\alpha}^{i} y^{\alpha}_{i} -
\bar f\, \bar y^{\dot\alpha i} \bar\pi_{\dot\alpha i}\right), \\[6pt]
S_r &=& c\, \epsilon_{rpq} \,y^{\alpha\dot{\beta}}_{p}
u_{q\,\alpha\dot{\beta}}+{\textstyle\frac{i}{2}}\left[f\, \pi_{\alpha}^{i}(\sigma_r)_{i}{}^{j}y^{\alpha}_{j}
- \bar f\, \bar y^{\dot\alpha i} (\sigma_r)_{i}{}^{j}\bar\pi_{\dot\alpha j}\right] \ ,
\end{array}
\end{equation}
where $u_{r\,\alpha\dot\beta}$ is given by formula \eqref{3.3}.
The independence of the first expression
in \eqref{3.20} on the parameter $c$ follows from the reality of the four-vector
coordinates $y^{\alpha\dot{\beta}}_{a} \sim (x^\mu, y^{\mu}_{r})$.

To describe the phase space structure of the model
\eqref{3.13} we calculate the momenta
$p_{\alpha\dot{\beta}}^{a}$, $p_{(\pi)}{}^{\alpha}_i$,
$p_{(\pi)}{}^{\dot{\alpha}i}$, $p_{(y)}{}_{\alpha}^i$,
$p_{(y)}{}_{\dot{\alpha}i}$ conjugate to $y^{\alpha\dot{\beta}}_{a}$,
$\pi_\alpha^i$, $\bar\pi_{\dot\alpha i}$, $y^{\alpha}_i$,
$\bar y^{\dot{\alpha}i}$. This leads to
the constraints (we set $c=f=1$ for simplicity)
\begin{equation}
\label{3.20a}
T_{\alpha\dot{\beta}}^{a} = p_{\alpha\dot{\beta}}^{a} - u_{\alpha\dot{\beta}}^{a} \approx 0 \,,
\end{equation}
\begin{equation}
\label{3.20b}
G_{\alpha}^i = p_{(y)}{}_{\alpha}^i - \pi_{\alpha}^i \approx 0 \,,\qquad
\bar G_{\dot\alpha i} = \bar p_{(y)}{}_{\dot\alpha i} - \bar\pi_{\dot\alpha i} \approx 0 \,,
\end{equation}
\begin{equation}
\label{3.20c}
F^{\alpha}_i = p_{(\pi)}{}^{\alpha}_i \approx 0 \,,\qquad
\bar F^{\dot\alpha i} = \bar p_{(\pi)}{}^{\dot\alpha i} \approx 0 \,.
\end{equation}
The remaining two (mass) constraints are given by \eqref{1.10}.

The constraints \eqref{3.20b} and \eqref{3.20c} are of second class.
Introducing the corresponding Dirac brackets $\{ A, B \}\,\to\,\{ A, B \}_{\ast}$
and eliminating by \eqref{3.20b} the momenta
$p_{(y)}{}_{\alpha}^i$, $\bar p_{(y)}{}_{\dot\alpha i}$ we get the
following set of Dirac brackets taking the canonical form
\begin{equation}
\label{3.21}
\{ y^{\gamma\dot{\delta}}_{a} , p_{\alpha\dot{\beta}}^{b} \}_{\ast}
= \delta_a^b \delta^\gamma_\alpha
\delta^{\dot{\delta}}_{\dot{\beta}} \, ,\qquad
\{ y^\alpha_i  , \pi^j_\beta\}_{\ast} = \delta_i^j \delta_\beta^\alpha \, ,
\qquad
\{ \bar{y}^{\dot{\alpha}i}  , \bar{\pi}_{\dot{\beta}j} \}_{\ast} =
\delta^i_j \delta_{\dot{\beta}}^{\dot{\alpha}}\, .
\end{equation}
The constraints \eqref{1.10} and \eqref{3.20a} are first class.
By quantizing the brackets \eqref{3.21} and introducing the realization
\begin{equation}
\label{3.22}
\hat y_a^{\alpha\dot{\beta}}=y_a^{\alpha\dot{\beta}}\,,\qquad
\hat{p}_{\alpha\dot{\beta}}^{\,a} = -i \frac{\partial}{\partial
y^{\alpha\dot{\beta}}_{a}}\ ,
\end{equation}
we obtain the $D{=}4$ unfolded equation
for the wave function
$ \Psi(y^{\alpha\dot{\beta}}_{a}, \pi^i_\alpha, \bar{\pi}_{\dot{\alpha}i})$:
\begin{equation}
\label{3.23}
\left( i \frac{\partial}{\partial y^{\alpha\dot{\beta}}_{a}}
+\pi_\alpha^i (\sigma^a)_{i}{}^{j} \bar{\pi}_{\dot{\beta}j}
\right)\Psi(y^{\alpha\dot{\beta}}_{a}, \pi^i_\alpha,
\bar{\pi}_{\dot{\alpha}i}) = 0 \; .
\end{equation}
The equation \eqref{3.23} has the solution ($a=0,1,2,3$)
\begin{equation}
\label{3.24}
\Psi(y^{\alpha\dot{\beta}}_{a}, \pi^i_\alpha,
\bar{\pi}_{\dot{\alpha}i}) = \exp \Big\{i \pi^i_\alpha (\sigma^a)_{i}{}^{j}
\bar{\pi}_{\dot{\beta}j}y^{\alpha\dot{\beta}}_{a}\Big\}\,
\psi(\pi^i_\alpha, \bar{\pi}_{\dot{\alpha}i})\ ,
\end{equation}
where the reduced wave functions $\psi(\pi, \bar{\pi})$ depend on complex $D{=}4$
spinorial momenta satisfying the mass constraints in \eqref{1.10}. For the general
model \eqref{3.14} ($c{\neq} 0$, $f{\neq} 0$) it follows from \eqref{3.20}
that all four variables $S^a$ are dynamical and that the reduced wave
function $\psi(\pi, \bar{\pi})$ does not satisfy  any
further constraints besides \eqref{1.10}.

The spinors $e_\alpha^i = (M)^{-\frac{1}{2}} \pi_\alpha^i$
($\bar{e}_{\dot{\alpha}i} = (\bar{M})^{-\frac{1}{2}}
\bar{\pi}_{\dot{\alpha}i}$) define a complex-holomorphic (complex
anti-holomorphic) spinorial $SL(2;\mathbb{C})$ Lorentz
frame ($SL(2;\mathbb{C})$ spinorial harmonics),
\begin{equation}
\label{3.25}
e^i_\alpha e^{\alpha j}= \epsilon^{ij} \ ,\quad  e^i_\alpha e_{\beta
i}= \epsilon_{\alpha\beta}\ ;\qquad \bar{e}^i_{\dot{\alpha}}
\bar{e}^{\dot{\alpha} j}= \epsilon^{ij} \ ,\quad
\bar{e}^i_{\dot{\alpha}} \bar{e}_{\dot{\beta}
i}= \epsilon_{\dot{\alpha}\dot{\beta}} \;,
\end{equation}
and the reduced wave function $\psi(\pi^i_\alpha, \bar{\pi}_{\dot{\alpha}i})$
in \eqref{3.24} depends
on an arbitrary element of the $SL(2;\mathbb{C})$  group (see also
\cite{BMSS83}). The six unconstrained degrees
of freedom can be described by the spinorial frame
$e_\alpha^i$ ($i$=1,2, $\alpha$=1,2, eq.~\eqref{3.25})
or by the vectorial frame given by the four-vectors $e_{\mu}{}^{a}$
($a=0,1,2,3$; $\mu=0,1,2,3$) satisfying the orthonormality relation \eqref{3.7}.
In particular, following \cite{BBTD88}, one can incorporate five
degrees of freedom into the pair of four-vectors
\begin{equation}
\label{3.27}
p_\mu^{(0)} =m e^{(0)}_\mu \equiv  p_\mu \ ,\qquad p_\mu^{(1)}=m e^{(1)}_\mu \equiv  q_\mu\ ,
\end{equation}
satisfying the conditions
\begin{equation}
\label{3.28}
p_\mu p^\mu =m^2 \, ,\qquad q_\mu q^\mu =-m^2\ ,\qquad p_\mu q^\mu =0\ .
\end{equation}
The four-vector $q_\mu$ parametrizes the sphere $\mathbb{S}^2$ in an
arbitrary Lorentz frame. The remaining sixth degree of freedom can be described by the
$SO(2)$ angle $0{\leq}\gamma{<}2\pi$, defined by the third vector $r_\mu$
\begin{equation}
\label{3.29}
p^{(2)}_\mu = m e^{(2)}_\mu = r_\mu \ ,\qquad r_\mu r^\mu =-m^2 \ ,\qquad p_\mu r^\mu = q_\mu r^\mu =0 \ .
\end{equation}
In the rest frame, $p_\mu=(m,0,0,0)$, the four-vector $r_\mu$ can
be parametrized as
\begin{equation}
\label{3.30}
r_\mu = (0,0, m\cos \gamma , m\sin \gamma) \ .
\end{equation}
Therefore, the reduced wave function \eqref{3.24} incorporating the mass constraints
\eqref{1.10} can be parametrized as
\begin{equation}
\label{3.31}
\psi (\pi^i_\alpha , \bar{\pi}_{\dot{\alpha}i})
\equiv \hat\psi(p_\mu, \mathbb{S}^2, \mathbb{S}^1)|_{p^2=m^2}\ ,
\end{equation}
where $\mathbb{S}^2$ is described by $q_\mu$ and $\mathbb{S}^1$ parametrizes $r_\mu$
by eq.~\eqref{3.30}. To describe the $D{=}4$ integer spin states
we may neglect the dependence on the $\mathbb{S}^1$ parameter; however for
half-integer spins the dependence on the angle $\gamma$
becomes necessary (see {\it e.g.} \cite{BBTD88}).

Let us consider now the model \eqref{3.14} for $f{=}\,0$, $c\,{\neq}\,0$, {\it i.e.} without
kinetic spinorial terms introduced by Vasiliev \cite{V+} in order to obtain
the unfolded equations (see \eqref{3.23}). We shall follow the arguments given for $D{=}3$ in
the last part of Sec.~\ref{gen}. When $f{=}\,0$ one obtains the constraints
\eqref{1.10}, \eqref{3.20a}, \eqref{3.20c} but not the constraints \eqref{3.20b}.
Using \eqref{3.7},  we introduce for \eqref{3.20a} and \eqref{3.20c}
the equivalent set of sixteen real and four pairs of complex-conjugated
constraints ($a,b\,{=}\,0,1,2,3$)
\begin{equation}
\label{3.37}
T^a_b = T_{\alpha\dot{\beta}}^{a} u^{\alpha\dot{\beta}}_{b}=
p_{\alpha\dot{\beta}}^{a} u^{\alpha\dot{\beta}}_{b}- m^2 \delta^a_b \approx 0\,,
\end{equation}
\begin{equation}
\label{3.37a}
F_a = {\textstyle\frac{1}{2}}\, \pi^i_\alpha (\sigma_ a)_i{}^j
p_{(\pi) j}^\alpha\approx 0 \ ,\qquad
\bar{F}_a = {\textstyle\frac{1}{2}}\, p_{(\bar{\pi})}^{\dot{\alpha}i}
(\sigma_ a)_i{}^j\bar{\pi}_{\dot{\alpha}j}
\approx 0 \ .
\end{equation}
The nonvanishing Dirac
brackets (see \eqref{3.21}) of the constraints
\eqref{3.37}, \eqref{3.37a} and \eqref{1.10} are ($F_a=(F_0,F_r)$)
\begin{equation}
\label{3.38a}
\{ F_q, F_r \}_\ast =-i\epsilon_{qrs}F_s  \,,\qquad
\{ \bar{F}_q, \bar{F}_r \}_\ast = i\epsilon_{qrs}\bar F_s \,,
\end{equation}
\begin{equation}
\label{3.38b}
\{ F_0, T^a_b\}_\ast = -{\textstyle\frac{1}{2}}\,T^a_b- {\textstyle\frac{1}{2}}\,\delta^a_b m^2\,, \qquad
\{ \bar{F}_0, T^a_b\}_\ast = -{\textstyle\frac{1}{2}}\,T^a_b- {\textstyle\frac{1}{2}}\,\delta^a_b m^2\,,
\end{equation}
\begin{equation}
\label{3.38c}
\{ F_r, T^a_0\}_\ast = -{\textstyle\frac{1}{2}}\,T^a_r- {\textstyle\frac{1}{2}}\,\delta^a_r m^2 \,, \qquad
\{ \bar{F}_r, T^b_0\}_\ast = -{\textstyle\frac{1}{2}}\,T^b_r- {\textstyle\frac{1}{2}}\,\delta^b_r m^2 \,,
\end{equation}
\begin{equation}
\nonumber
\{ F_r, T^a_q\}_\ast = -{\textstyle\frac{i}{2}}\,\epsilon_{rqs}T^a_s- {\textstyle\frac{i}{2}}\,\epsilon_{rqs}\delta^a_s m^2
-{\textstyle\frac{1}{2}}T^a_0 \delta_{rq} - {\textstyle\frac{1}{2}}m^2 \delta^a_0\delta_{rq} \,,
\end{equation}
\begin{equation}
\label{3.38d}
\{ \bar{F}_r, T^a_q\}_\ast = {\textstyle\frac{i}{2}}\,\epsilon_{rqs}T^a_s+ {\textstyle\frac{i}{2}}\,\epsilon_{rqs}\delta^a_s m^2
-{\textstyle\frac{1}{2}}T^a_0 \delta_{rq} - {\textstyle\frac{1}{2}}m^2 \delta^a_0\delta_{rq} \,,
\end{equation}
\begin{equation}
\label{3.38e}
\{ F_0, \mathcal{M}\}_\ast = - \mathcal{M}+2M\,, \qquad
\{ \bar{F}_0, \bar{\mathcal{M}}\}_\ast= -\bar{\mathcal{M}}+2\bar{M}\ ,
\end{equation}
where $q,r,s$=1,2,3. We see that the $8$ real constraints $F_a$, $\bar{F}_a$
provide a partial gauge fixing of the $18$=16+2 gauge transformations, which
in the case $f{\neq}\, 0$ are generated by the 18=16+2 first class constraints $T^a_b$,
$\mathcal{M}$, $\bar{\mathcal{M}}$. One can calculate that if $c\,{\neq}\, 0$
the variants $f{\neq}\, 0$ and $f{=}0$ of the model \eqref{3.14}
have the same number of twelve real physical (non-gauge) degrees of freedom
but different number (18 for $f{\neq}\, 0$ and 10 for $f{=}0$) of local
({\it i.e.} $\tau$-dependent) gauge parameters.

We add that for a $D{=}\,4$ particle of mass $m$ and fixed spin $s$
the physical phase space has eight degrees of freedom, with the spin
degrees represented {\it e.g.} by the coordinates on the sphere $\mathbb{S}^2$
\cite{Schul68,BBTD88}. In such a theory the relation \eqref{3.12} that
determines the fixed spin value $s$ is a first class constraint.
If this constraint is removed, the resulting theory with arbitrary spin
$s$ has then ten degrees of freedom. It will be shown in Sec.\,3.4 that
the wave function solving the model \eqref{3.14} describes twelve degrees
of freedom due to the multiplicity that is associated with each
value of the different spins. We shall reduce the twelve degrees of freedom
to ten, as required by a HS theory with nondegenerate spin spectrum,
by imposing an harmonicity constraint (see \eqref{cond-wv} below)
on the wave function.

\subsection{$D{=}\,4$ bitwistorial description of HS massive
multiplets}
\label{4desc}

Following the procedure in Sec.~\ref{D=3} for $D$=3, we
now express the action \eqref{3.14} just in terms of a pair of
$D{=}\,4$ twistor coordinates (eq.~\eqref{3.1}) by postulating the
incidence relations \eqref{3.19}. With $f{\neq} 0$ ($c$ may be
arbitrary) this leads to the following two-twistorial action
with two complex-conjugated Lagrange multipliers $\mu,\bar{\mu}$
\begin{equation}
\label{3.39}
\tilde{S}^{(4)} = \int d \tau \left[ \pi^i_\alpha\dot\omega_i^\alpha
+ \mu \left(\pi^i_\alpha \pi^\alpha_i +2M\right) + \textrm{h.c.}
\right] \ .
\end{equation}
The model \eqref{3.39} contains only two complex-conjugated
spinorial mass constraints \eqref{1.10}. When $f=0$ and $c\neq 0$,
as it follows from formulae \eqref{3.20}, one still has to to impose
one additional constraint via a Lagrange multiplier
\begin{equation}
\label{3.40}
S_0 = -{\textstyle\frac{i}{2}}\,\bar{Z}_{Ai} {Z}^{Ai} \approx 0\, .
\end{equation}

In order to find the first and second class constraints
we use the canonical PB that follow from the $\tilde{S}^{(4)}$
action \eqref{3.39}
\begin{equation}
\label{3.41}
\{ \bar{Z}_{Ai},{Z}^{Bj}   \} = \delta_{i}^{j} \delta^B_A \ .
\end{equation}
One can check that
$\{ \mathcal{M}, S^0 \} \neq 0$, $\{ \bar{\mathcal{M}}, S^0  \} \neq 0$.
Further, we replace the two complex-conjugated constraints $\mathcal{M}$,
$\bar{\mathcal{M}}$ by a pair of real constraints
\begin{equation}\label{3.43}
\begin{array}{rcl}
\phi_1 &=& \frac{1}{2}\,(\mathcal{M}+\bar{\mathcal{M}}) =
\frac{1}{2}\,( \pi^i_\alpha \pi^\alpha_i + \textrm{h.c.}) -
M_1=0 \ ,  \\ [5pt]
\phi_2 &=& \frac{i}{2}\,(\mathcal{M}- \bar{\mathcal{M}}) =
\frac{i}{2}\,( \pi^i_\alpha \pi^\alpha_i - \textrm{h.c.}) -
M_2=0 \ ,
\end{array}
\end{equation}
where $M= M_1+iM_2$. The PB of the constraints ($S_0$, $\phi_1$,
$\phi_2$) are
\begin{equation}\label{3.44}
\begin{array}{rcl}
\{ S_0,\phi_1 \} &=& \phi_2 + M_2\,, \\ [5pt]
\{ S_0,\phi_2 \} &=& -\phi_1-M_1\,, \\ [5pt]
\{ \phi_1, \phi_2 \} &=& 0\ .
\end{array}
\end{equation}
The PBs in eq,~\eqref{3.44} show that the generators $S_0$,
${\phi}^\prime_1 = \phi_1 +M_1$, ${\phi}^\prime_2 = \phi_2 +M_2$
describe an $E(2)$ algebra, $\{ S_0,\phi^\prime_1 \} = \phi^\prime_2\ ,\;
\{ S_0,\phi^\prime_2 \} = -\phi^\prime_1 \ ,\;
\{ \phi^\prime_1, \phi^\prime_2 \} = 0$.

The shifts ${\phi}^\prime_1,{\phi}^\prime_2\,\to\,{\phi}_1,{\phi}_2$ of
the generators of the translation sector of $E(2)$ may be considered
as producing spontaneously broken symmetries. Indeed, after quantization of PB  \eqref{3.44}
one can consider that the action of the $E(2)$ generators
($\hat S_0,\hat{\phi}^\prime_1,\hat{\phi}\prime_2$) annihilates the vacuum $|\,0\rangle$.
Then, the quantized relations \eqref{3.44} are consistent only if $\hat S_0|\,0\rangle=0$,
$\hat{\phi}^\prime_{1,2}|\,0\rangle=0\, \Rightarrow \hat{\phi}_{1,2}|\,0\rangle=M_{1,2}|\,0\rangle\neq0$.
This means that if we look at $\hat{\phi}_{1}$, $\hat{\phi}_{2}$ as generating the
two translational symmetries of $E(2)$ these have to be spontaneously
broken\footnote{We recall that the symmetry associated with
a Lie algebra generator $\hat X$ is spontaneously broken if
$\hat X|\,0\rangle \neq0$ \cite{Orz}. The phenomenon above described is that
if $\hat{\phi}_{1}$, $\hat{\phi}_{2}$ are considered as translation generators,
then we cannot longer ignore that the true algebra is larger and that,
in it, the constants determine a central subalgebra. Taking a
basis that it is not a subalgebra led to the symmetry breaking  above.}.
Similarly, if we introduce another choice of generators
\begin{equation}
\label{3.45}
\widetilde{\phi}_1 = M_1 \phi_1 +M_2 \phi_2\ ,\qquad
\widetilde{\phi}_2 = M_2 \phi_1 -M_1 \phi_2\ ,
\end{equation}
the PB \eqref{3.44} will be rewritten as representing $E(2)$ algebra broken spontaneously only in
one translational direction generated by $\widetilde{\phi}_1$
\begin{equation}
\label{3.46}
\begin{array}{rcl}
\{ S_0,\widetilde{\phi}_1 \} &=& \widetilde{\phi}_2\,,  \\ [5pt]
\{ S_0,\widetilde{\phi}_2 \} &=& -\widetilde{\phi}_1-m^2\,, \\ [5pt]
\{ \widetilde{\phi}_1, \widetilde{\phi}_2 \} &=& 0\ ,
\end{array}
\end{equation}
where $m^2=|M|^2= M_1^2+M_2^2$. We see from \eqref{3.46} that the constraint
$\widetilde{\phi}_1 $ is of first class, and $\widetilde{\phi}_2$, $S_0$
form a pair of second class constraints.

It turns out nevertheless that the number of physical
phase space degrees of freedom is the same and equal to twelve, irrespectively
of the value of the parameter $f$. In fact,
\begin{enumerate}
\item
if $f\,{\neq}\, 0$ we have two first class
constraints \eqref{1.10}, {\it i.e.} in 16-dimensional two-twistor
phase space the number of degrees of freedom is $16-2\times 2=12$.
\item
if $f\,{=}\,0$ and $c\,{\neq}\, 0$ we get three constraints satisfying the PBs \eqref{3.46},
one first class and two second class. The count of degrees of
freedom is the same: $16-1\times 2- 2\times 1 =12$.
\item
If $f\,{=}\,0$ and $c\,{=}\,0$ we obtain the model of massive spinless particle
(see formulae \eqref{3.14}-\eqref{3.18}), with six-dimensional physical phase space.
\end{enumerate}

\noindent
In the fist two cases we obtain the twelve dimensions of physical phase
space by doubling the number of independent coordinates that parametrize
the six-dimensional manifold $SL(2;\mathbb{C})$; in accordance with
\eqref{3.31}, the reduced wave function is defined on this manifold.\\

To relate more closely our description with the spin degrees of freedom,
let us recall the Lorentz-invariant spin variables $S_r$ defined
by eq.~\eqref{3.18}. Using the PB relations in \eqref{3.41},
one can show that the bilinears $S_r$ satisfy the
$so(3)\simeq su(2)$ PB algebra  ($q,p,r=1,2,3$)
\begin{equation}
\label{3.47}
    \{ S_q, S_p\} = \epsilon_{qpr} S_r\ .
\end{equation}
In particular, if $S_r{\approx}\,0 \Rightarrow W_{\alpha
\dot{\beta}}{\approx}\ 0$ (see \eqref{3.10}), {\it i.e.}
the spin is equal to zero. In our twistorial
model $S_r\neq 0$ (see \eqref{3.20}) and after quantization ($S_r
\rightarrow \hat{S}_r$)) we obtain from \eqref{3.47} the $so(3)$ algebra of
Lorentz-invariant spin generators $\hat{S}_r$
\begin{equation}\label{3.48}
[ \hat{S}_q, \hat{S}_p \,] = i\epsilon_{qpr} \hat{S}_r\ .
\end{equation}
The mass shell constraints, after using the bitwistor formula \eqref{3.4} for
the four-momentum, provide the generalized Dirac equation
with complex mass $M$ and four-components complex Dirac spinors
\begin{equation}
\label{3.49}
\pi^{\beta i} p_{\alpha\dot{\beta}} = M \bar{\pi}_{\dot{\alpha}}^i
  \ ,\qquad
  p_{\alpha\dot{\beta}} \bar{\pi}^{\dot{\beta} i} = \bar{M}
  \pi_\alpha^i\ .
\end{equation}
Further, in our two-twistor framework we obtain as well the generalization of
eqs.~\eqref{3.49} for the set of three auxiliary fourmomenta ($r=1,2,3$),
\begin{equation}\label{3.50}
  \pi^{\alpha i} p_{\alpha\dot{\beta}}^{\,r} = M \bar{\pi}_{\dot{\beta}}^j(\sigma^r)_j{}^{i}
  \ ,\qquad
  p_{\alpha\dot{\beta}}^{\,r} \bar{\pi}^{\dot{\beta}j} = \bar{M}
  \pi_{\alpha}^{i} (\sigma^r)_i{}^{j}\ .
\end{equation}

  To replace complex value $M=\frac{1}{\sqrt{2}}\,e^{i\varphi}m$
by a real one $m$ let us observe that the action \eqref{3.14} is invariant under
the following global phase transformations
\begin{equation}\label{tr-ph}
\begin{array}{c}
\pi_{\alpha}^{\prime\, i}=e^{i\varphi/2}\pi_{\alpha}^{i}\,,\qquad
\bar\pi_{\dot\alpha i}^{\prime}=e^{-i\varphi/2}\bar\pi_{\dot\alpha i}\, ,\\ [6pt]
f^{\prime}=e^{-i\varphi/2}f\,,\qquad
\bar f^{\prime}=e^{i\varphi/2}\bar f\,;\qquad
\rho^{\prime}=e^{-i\varphi}\rho\,,\qquad
\bar \rho^{\prime}=e^{i\varphi}\bar \rho \,,
\end{array}
\end{equation}
where $e^{2i\varphi}=M/\bar M$. The $D{=}\,4$ mass constraints \eqref{1.10} are expressed
in terms of $\pi_{\alpha}^{\prime\, i}$, $\bar\pi_{\dot\alpha i}^{\prime}$ by
({\it cf.} eq.~\eqref{2.4}  for $D$=3)
\begin{equation}
\label{1.10-a}
\pi_{\alpha}^{\prime\, i} \pi^{\prime\,\alpha}_i +\sqrt{2}\,m =0\,,\qquad
\bar{\pi}_{\dot{\alpha}}^{\prime\, i}
\bar{\pi}^{\prime\, \dot{\alpha}}_i +\sqrt{2}\,m=0 \, .
\end{equation}
For the Weyl spinors $\pi_{\alpha}^{\prime\, i}$, $\bar\pi_{\dot\alpha i}^{\prime}$
we get the equations \eqref{3.49}, \eqref{3.50} with $M$ replaced by $m$.
The transformations \eqref{tr-ph} do not affect the $SL(2;\mathbb{C})$ part
of the variables $\pi_{\alpha}^{i}$
(see next section)
because they change only the determinant of $2{\times}2$ matrix $\pi_{\alpha}^{\, i}$,
which is parametrized by the coset $GL(2;\mathbb{C})/SL(2;\mathbb{C}) \simeq  GL(1;\mathbb{C})$,
parametrized by an arbitrary complex mass parameter.

\subsection{$D{=}\,4$ bitwistor wave function of HS massive
multiplet}\label{4wf}

Our $D{=}\,4$ dynamical bitwistorial system is described by twistorial
coordinates see \eqref{3.1}) in terms of the
variables $\pi^j_{\alpha}$, $\bar\pi_{\dot\alpha k}$, $\omega^{\alpha}_{k}$,
$\bar\omega^{\dot\alpha i}$ endowed with the canonical PBs
\begin{equation}
\label{PB-tw-s}
\{ \omega^{\alpha}_{i}, \pi^j_{\beta} \} =\delta^{\alpha}_{\beta}\delta_i^j\,,\qquad
\{ \bar\omega^{\dot\alpha i} ,\bar\pi_{\dot\beta j}\} =\delta^{\dot\alpha}_{\dot\beta}\delta^i_j\,,
\end{equation}
constrained by the mass constraints
$\mathcal{M}$, $\bar{\mathcal{M}}$ (eqs.~ (\ref{1.10})).
Further we shall assume $f\,{=}\,0$ and $c\,{\neq}\, 0$.
In such a case we should add the constraint (\ref{3.40})
\begin{equation}
\label{S0-tw}
V=-2S_0=i\left(\pi^i_{\alpha}\bar\omega^{\alpha}_{i}-
\bar\pi_{\dot\alpha i}\omega^{\dot\alpha i}\right)\approx0
\end{equation}
with nonvanishing PBs
\begin{equation}
\label{PB-constr-s}
\{ V ,\mathcal{M}\} =2i\mathcal{M}+4iM\quad ,
\quad \{ V ,\bar{\mathcal{M}}\} =-2i\bar{\mathcal{M}}-4i\bar M\,.
\end{equation}

   The constraints $\mathcal{M}$, $\bar{\mathcal{M}}$ can be equivalently described by
\begin{equation}
\label{F-s1}
F_1= \bar M\mathcal{M}+M\bar{\mathcal{M}}\,,\quad F_2=i(\bar M\mathcal{M}-M\bar{\mathcal{M}}) \, ,
\end{equation}
One can check easily that the constraints $V$ and $F_2$ are second class.
For the local gauge transformations generated by the constraint $F_1$
we introduce the gauge fixing condition
\begin{equation}\label{gauge}
G=\pi^i_{\alpha}\bar\omega^{\alpha}_{i}+ \bar\pi_{\dot\alpha i}\omega^{\dot\alpha i}\approx0\,,
\end{equation}
described by the generator of scale transformations (dilatations) for twistorial variables.
Further, using \eqref{PB-tw-s}, \eqref{1.10} and \eqref{gauge} one obtains
\begin{equation}\label{PB-G-s1}
\{ G ,\mathcal{M}\}=2\mathcal{M}+4M\,,\quad \{ G ,\bar{\mathcal{M}}\}=2\bar{\mathcal{M}}+4\bar M\,.
\end{equation}
The PB of the constraints $V$, $G$, $F_1$ and $F_2$ are
\begin{equation}\label{PB-VGF}
\begin{array}{l}
\{ G ,F_1\}=2F_1+8M\bar M\,,\qquad \{ G ,F_2\}=2F_2\,, \\ [5pt]
\{ V ,F_1\}=2F_2\,,\qquad \{ V ,F_2\}=-2F_1-8M\bar M\,.
\end{array}
\end{equation}
Then, the Dirac brackets (DB) that account for the four second class constraints \eqref{PB-VGF}
are defined by the formula
\begin{equation}
\label{DB}
\begin{array}{c}
\{ A,B \}_{*}=\{ A,B \}+ \\[5pt]
{\textstyle\frac{1}{8M\bar M}}\Big[
\{ A,G \}\{ F_1,B \}-\{ A,F_1 \}\{ G,B \}
-\{ A,V \}\{ F_2,B \}+\{ A,F_2 \}\{ V,B \}
\Big]\,.
\end{array}
\end{equation}
This gives for the twistor components the DBs
\begin{equation}
\label{DB-tw-l}
\{ \pi^k_{\alpha}, \pi^j_{\beta} \}_{*}=
\{ \bar\pi_{\dot\alpha k} ,\bar\pi_{\dot\beta j}\}_{*}=
\{ \pi^k_{\alpha} ,\bar\pi_{\dot\beta j}\}_{*}=0\,,
\end{equation}
\begin{equation}
\label{DB-tw-2}
\{ \omega^{\alpha}_{k}, \pi^j_{\beta} \}_{*}=\delta^{\alpha}_{\beta}\delta_k^j+
{\textstyle\frac{1}{2M}}\pi^{\alpha}_{k}\pi^j_{\beta}\,,\qquad
\{ \bar\omega^{\dot\alpha k} ,\bar\pi_{\dot\beta j}\}_{*}=\delta^{\dot\alpha}_{\dot\beta}\delta^k_j-
{\textstyle\frac{1}{2\bar M}}\bar\pi^{\dot\alpha k} \bar\pi_{\dot\beta j}\,,
\end{equation}
\begin{equation}
\label{DB-tw-3}
\{  \bar\omega^{\dot\alpha k}, \pi^j_{\beta} \}_{*}=0\,,\qquad
\{ \omega^{\alpha}_{k},\bar\pi_{\dot\beta j}\}_{*}=0\,,
\end{equation}
\begin{equation}
\label{DB-tw-4}
\{ \omega^{\alpha}_{k}, \omega^{\beta}_{j} \}_{*}=-{\textstyle\frac{1}{M}}
\left(\pi^{\alpha}_{k}\bar\omega^{\beta}_{j}-\pi_j^{\beta} \bar\omega^{\alpha}_{k}\right)\,,\qquad
\{ \bar\omega^{\dot\alpha k} ,\bar\omega^{\dot\beta j}\}_{*}=
{\textstyle\frac{1}{\bar M}}\left(\bar\pi^{\dot\alpha k}\omega^{\dot\beta j}
- \bar\pi^{\dot\beta j}\omega^{\dot\alpha k}\right),
\end{equation}
\begin{equation}
\label{DB-tw-5}
\{ \omega^{\alpha}_{k}, \bar\omega^{\dot\beta j}\}_{*}=0\,.
\end{equation}

Below we will consider the $(\pi,\bar\pi)$-realization of quantized
version of the DB algebra (\ref{DB-tw-l})-(\ref{DB-tw-5}).
In such a realization, after using the ordering with $\pi$'s
at the left and $\omega$'s at the right, we obtain
$\hat\pi_{\alpha}^{k}=\pi_{\alpha}^{k}$,
$\hat{\bar\pi}_{\dot\alpha k}=\bar\pi_{\dot\alpha k}$ and
\begin{equation}
\label{op-om}
\hat{\omega}{}^{\alpha}_{k}=i\frac{\partial}{\partial\pi_{\alpha}^{k}}+
\frac{i}{2M}\,\pi^{\alpha}_{k}\,\pi_{\beta}^{j} \frac{\partial}{\partial\pi_{\beta}^{j}}\,,\qquad
\hat{\bar\omega}{}^{\dot\alpha k}=i\frac{\partial}{\partial\bar\pi_{\dot\alpha k}}-
\frac{i}{2\bar M}\,\bar\pi^{\dot\alpha k}\,\bar\pi_{\dot\beta j} \frac{\partial}{\partial\bar\pi_{\dot\beta j}}\,.
\end{equation}
one checks that in the presence of $D{=}\,4$ mass constraints \eqref{1.10}
the constraints (\ref{S0-tw}), (\ref{gauge}) are satisfied in the strong sense:
$\hat\pi_{\alpha}^{k}\hat\omega^{\alpha}_{k}\equiv0$,
$\hat{\bar\pi}_{\dot\alpha k}\hat{\bar\omega}{}^{\dot\alpha k}\equiv0$.

Taking into account the expressions (\ref{op-om}) we obtain
the quantum counterparts of the quantities (\ref{Sr-s}) as the spin operators
\begin{equation}\label{S-qu}
\hat S_r={\frac{1}{2}}\left(\pi^i_{\alpha}\frac{\partial}{\partial\pi_{\alpha}^{k}}-
\bar\pi_{\dot\alpha k}\frac{\partial}{\partial\bar\pi_{\dot\alpha i}}\right)(\sigma_r)_i{}^k\,.
\end{equation}
Using (\ref{3.12}), the square of the Pauli-Luba\'nski vector becomes
$\hat W^\mu \hat W_\mu=-m^2\hat S^r\hat S^r$, which will be used later to define spin states.

Thus, the twistorial wave function is defined on the space parametrized by
$\pi_{\alpha}^{i}$, $\bar\pi_{\dot\alpha i}$ which satisfy the constraints
$\mathcal{M}$, $\bar{\mathcal{M}}$ (eq.~(\ref{1.10})),
and the matrix
\begin{equation}
\label{sl2c}
g_{\alpha}{}^{i}=M^{-1/2}\pi_{\alpha}^{i}
\end{equation}
defines the $SL(2,\mathbb{C})$ group manifold.~Thus, the twistorial wave function
is defined on $SL(2,\mathbb{C})$ parametrized by $\pi_{\alpha}^{i}$,
so that $\Psi=\Psi(\pi_{\alpha}^{i},\bar\pi_{\dot\alpha i})$.
One can use the well known decomposition of $SL(2,\mathbb{C})$ elements
\begin{equation}
\label{sl2c-exp}
g=h\,v\,,\qquad g_{\alpha}{}^{i}=h_{\alpha}{}^{\tt k}v_{\tt k}{}^{i} \, ,
\end{equation}
in terms of the product of an hermitian matrix $h=h^\dagger$ with unit determinant and
an $SU(2)$ matrix $v$, $v^\dagger v=1$
(in the above formulae, the $v_{\tt k}{}^{i}$ play the
role of $\psi$ in \eqref{su11-par} for $D$=3).
The three parameters of the matrix $h$ describe four-momenta on the mass shell,
and the three parameters of the matrix $v$ correspond to the spin algebra \eqref{3.48}.
The matrix $h$ parametrizes the coset $SL(2,\mathbb{C})/SU(2)$ which defines
the three-dimensional mass hyperboloid for timelike four-momenta
which does not depend on the $v_{\tt k}{}^{i}$ variables
(as in $D$=3 eqs.~\eqref{mom-comp} do not depend on $\psi$).
So, the definition \eqref{3.4} can be rewritten as follows
\begin{equation}
\label{p-n}
p_{\alpha\dot{\beta}} =
h_\alpha{}^{\tt i} \bar{h}_{\dot{\beta}{\tt i}}\, ,
\end{equation}
where $\bar{h}_{\dot{\alpha}{\tt i}}=(h_{\alpha}{}^{\tt i})^*$ and $\alpha$=1,2 and ${\tt i}$=1,2.

The unitary matrix $v$ paramerizes $\mathbb{S}^3\sim SU(2)$ and is linked with
the spin degrees of a massive particle. In particular, the operators \eqref{S-qu}
expressed by the variables \eqref{sl2c-exp} take the form
\begin{equation}
\label{S-qu1}
\hat S_r={\textstyle\frac{1}{2}}\,(\sigma_r)_j{}^k\,v_{\tt i}{}^j\frac{\partial}{\partial v_{\tt i}{}^{k}}\,.
\end{equation}
We can consider the variables $v_{\tt i}{}^{k}$ as the harmonic variables that
were introduced early to describe $N{=}2$ superfield formulations
(see, for example, \cite{GIOS}). In particular, it is useful to introduce the notation
\begin{equation}
\label{harm}
v_{\tt i}{}^{k}=(v_{\tt i}{}^{1},v_{\tt i}{}^{2})=(v_{\tt i}^{+},v_{\tt i}^{-})\,,\qquad
v^{+ {\tt i}}v^-_{\tt i}=1\,,\qquad
(v^\pm_{\tt i})^*=\mp v^{\mp {\tt i}}\,.
\end{equation}
Then, the operators  \eqref{S-qu1} take the form
\begin{equation}
\label{S-qu1a}
D^0\equiv 2\hat S_3=v^+_{\tt i}\frac{\partial}{\partial v^+_{\tt i}}-v^-_{\tt i}\frac{\partial}{\partial v^-_{\tt i}}\,,
\qquad
D^{\pm\pm}\equiv \hat S_1\pm {\tt i}\hat S_2=v^\pm_{\tt i}\frac{\partial}{\partial v^\mp_{\tt i}}\,,
\end{equation}
and the square of the Pauli-Luba\'nski vector is given by the formula
\begin{equation}
\label{PL-harm}
\hat W^\mu \hat W_\mu =
-\frac{m^2}{4}\left[ \left(D^0\right)^2+2\left\{ D^{++}, D^{--} \right\} \right]\,.
\end{equation}

Since the variables $v^\pm_{\tt i}$ parametrize a compact space, the general wave
function on $SL(2,\mathbb{C})$ has the following
harmonic expansion (we use the $SU(2)$-covariant expansion from \cite{GIOS})
\begin{equation}
\label{wv-ncov}
\Psi(h_{\alpha}{}^{\tt i},v_{\tt i}{}^{k})=
\sum_{K,N=0}^\infty v^+_{({\tt i}_1}\ldots v^+_{{\tt i}_N}v^-_{{\tt j}_1}\ldots v^-_{{\tt j}_K)}\,
f^{{\tt i}_1\ldots {\tt i}_N{\tt j}_1\ldots {\tt j}_K}(h)\,,
\end{equation}
where the coefficient fields $f^{{\tt i}_1\ldots {\tt i}_N{\tt j}_1\ldots {\tt j}_K}(h)=
f^{({\tt i}_1\ldots {\tt i}_N{\tt j}_1\ldots {\tt j}_K)}(h)$
are symmetric with respect to all indices
because the antisymmetric contributions involving
factors in $v^+$ and $v^-$ disappear due to the formula
\begin{equation}
\label{norm-harm}
v^+_{{\tt i}}v^-_{{\tt j}}-v^+_{{\tt j}}v^-_{{\tt i}}=\epsilon_{{\tt i}{\tt j}}\,,
\end{equation}
which follows from the second expression in the
definition of harmonic variables \eqref{harm}. These coefficient
fields depend on the on-shell four-momenta due to \eqref{p-n},
$f^{{\tt i}_1\ldots {\tt i}_N{\tt j}_1\ldots {\tt j}_K}(h)=
f^{{\tt i}_1\ldots {\tt i}_N{\tt j}_1\ldots {\tt j}_K}(p_\mu)$.
Such functions defined on the mass hyperboloid can be expanded
into $SL(2;\mathbb{C})$ irreducible representations belonging
to the principal series of the first kind \cite{GelGV}.

Each monomial of the variables $v^\pm_{\tt i}$ in the expansion \eqref{wv-ncov}
is an eigenvector of the Casimir operator \eqref{PL-harm}:
\begin{equation}
\label{PL-eig-vect}
\begin{array}{c}
\hat W^\mu \hat W_\mu\, v^+_{{\tt i}_1}\ldots v^+_{{\tt i}_N}v^-_{{\tt j}_1}\ldots v^-_{{\tt j}_K}\,
f^{{\tt i}_1\ldots {\tt i}_N{\tt j}_1\ldots {\tt j}_K}= \hspace{6cm}\\[6pt]
\hspace{6cm} -m^2\,s(s+1)\,v^+_{{\tt i}_1}\ldots v^+_{{\tt i}_N}v^-_{{\tt j}_1}\ldots v^-_{{\tt j}_K}\,
f^{{\tt i}_1\ldots {\tt i}_N{\tt j}_1\ldots {\tt j}_K},
\end{array}
\end{equation}
where
$s=\frac{N+K}{2}$. So, the expression \eqref{wv-ncov} is in fact the
general expansion into arbitrary spin states.
By means of the nonsingular transformation $v\to g=hv$, {\it i.e.}
$v^\pm_{\tt i} \to \pi^\pm_{\alpha}$ or
$v^\mp_{\tt i} \to \bar\pi^\pm_{\dot\alpha}$
where
\begin{equation}
\label{pi-pm}
(\pi^+_{\alpha},\pi^-_{\alpha})=(\pi^1_{\alpha},\pi^2_{\alpha})\,,\qquad
(\bar\pi^+_{\dot\alpha},\bar\pi^-_{\dot\alpha})=(\bar\pi_{\dot\alpha 2},-\bar\pi_{\dot\alpha 1})\,,
\end{equation}
and by redefining the component fields, we can rewrite the expression \eqref{wv-ncov}
in $SL(2,\mathbb{C})$-covariant form, but we would like to stress that the
spin content in the expansion \eqref{wv-ncov} is degenerate.
This degeneracy can be however removed by the harmonic condition
on the wave function  (see also \cite{GIOS})
\begin{equation}
\label{cond-wv}
D^{++}\,\tilde\Psi^{(+)}=0\,.
\end{equation}
Since that the monomials $v^+_{({\tt i}_1}\ldots v^+_{{\tt i}_N}v^-_{{\tt j}_1}\ldots v^-_{{\tt j}_K)}$
form the basis, as a solution of \eqref{cond-wv}, we obtain the following wave function
\begin{equation}
\label{wv-ncov1}
\tilde\Psi^{(+)}(h_{\alpha}{}^{\tt i},v_{\tt i}^{\pm})=\sum_{N=0}^\infty v^+_{{\tt i}_1}\ldots v^+_{{\tt i}_N}\,
f^{{\tt i}_1\ldots {\tt i}_N}(h)\,.
\end{equation}
This twistor wave function rewritten in Lorentz covariant way takes the form
\begin{equation}
\label{wv-tw}
\tilde\Psi^{(+)}(\pi_{\alpha}^{\pm},\bar\pi_{\dot\alpha}^{\pm})=
\sum_{N=0}^\infty \pi^+_{\alpha_1}\ldots \pi^+_{\alpha_N}\,
\psi^{\alpha_1\ldots \alpha_N}(p_\mu)\,.
\end{equation}
Note that twistor wave function \eqref{wv-tw} also depends on $\pi_{\alpha}^{-}$ and
$\bar\pi_{\dot\alpha}^{\pm}$ through $p_\mu$ in the argument of the component fields.

Spin $s$=$L/2$ massive particles are described by the fields
$\psi^{\alpha_1\ldots \alpha_L}(p_\mu)$. The corresponding spacetime
fields are obtained by an integral Fourier-twistor transform which combines
the Fourier and twistor  transformations.
More explicitly, by means of these integral transformations we can obtain
the following multispinor fields, all with a total of $L$ undotted
plus dotted indices,
\begin{equation}
\label{wv-st}
\begin{array}{rcl}

\phi_{\alpha_1\ldots \alpha_L}(x)&=&
{\displaystyle \int}d^6\pi\,e^{-ix^\mu p_\mu} \pi^-_{\alpha_1}\ldots \pi^-_{\alpha_L}
\tilde\Psi^{(+)}(\pi^{\pm},\bar\pi^{\pm})\,,\\ [7pt]

\phi_{\alpha_1\ldots \alpha_{L-1}}{}^{\dot\beta_1}(x) &=&
{\displaystyle \int}d^6\pi\,e^{-ix^\mu p_\mu} \pi^-_{\alpha_1}\ldots
 \pi^-_{\alpha_{L-1}}\bar\pi^{-\dot\beta_1}
\tilde\Psi^{(+)}(\pi^\pm,\bar\pi^\pm)\,,\\ [7pt]

\phi_{\alpha_1\ldots \alpha_{L-2}}{}^{\dot\beta_1 \dot\beta_2}(x) &=&
{\displaystyle \int}d^6\pi\,e^{-ix^\mu p_\mu} \pi^-_{\alpha_1}\ldots
\pi^-_{\alpha_{L-2}}\bar\pi^-{}^{\dot\beta_1}\bar\pi^-{}^{\dot\beta_2}
\tilde\Psi^{(+)}(\pi^{\pm},\bar\pi^{\pm})\,,\\ [7pt]

&&\quad ................................. \\ [7pt]
\phi^{\dot\beta_1\ldots \dot\beta_L}(x)&=&
{\displaystyle \int}d^6\pi\,e^{-ix^\mu p_\mu} \bar\pi^{-\dot\beta_1}\ldots \bar\pi^{-\dot\beta_L}
\tilde\Psi^{(+)}(\pi^{\pm},\bar\pi^{\pm})
\end{array}
\end{equation}
where $p_\mu$ is defined by \eqref{3.4} as a bilinear product of twistors.
In the integrals \eqref{wv-st} for a given $L$,
only the term $\pi^+_{\alpha_1}\ldots \pi^+_{\alpha_L}\,\psi^{\alpha_1\ldots \alpha_L}(p_\mu)$
in the twistorial wave function \eqref{wv-tw},
with $U(1)$ harmonic charge $q=L$ (see \cite{GIOS}), gives a non-zero contribution.
Note that $d^6\pi=d^3 h\, d^3 v$ where $d^3 v$ is the harmonic measure on $SU(2)$
manifold whereas the measure on the Lobachevski mass hyperboloid $d^3 h{=}d\Omega$
may be written as $d\Omega=d^3\vec{p}/(2p_0)$ after using
the the relation \eqref{p-n}. Since all fields in eqs.~\eqref{wv-st} are
derived from eq.~\eqref{wv-tw}, they have to be related.
Denoting by $N (M)$ the total number of undotted (dotted) indices, $L$=$N$+$M$,
we can show that the multispinors $\phi_{\alpha_1\ldots \alpha_N}{}^{\dot\beta_1\ldots \dot\beta_{M}}$
($=\phi_{(M,N)}$ for short) in \eqref{wv-st}, symmetric in both the $\alpha$ and the $\dot\beta$
indices, satisfy the following sequence of Dirac-Fierz-Pauli field equations
\begin{equation}
\label{DFP-eq}
\begin{array}{rcl}
i\partial_{\alpha \dot\beta_M}\phi_{\alpha_1\ldots \alpha_N}{}^{\dot\beta_1\ldots \dot\beta_{M}}&=&
m\,\phi_{\alpha \alpha_1\ldots \alpha_N}{}^{\dot\beta_1\ldots \dot\beta_{M-1}}\,, \\ [7pt]

i\partial^{\dot\beta_M \alpha }\phi_{\alpha\alpha_1\ldots \alpha_N}{}^{\dot\beta_1\ldots \dot\beta_{M-1}}&=&
m\,\phi_{\alpha_1\ldots \alpha_N}{}^{\dot\beta_1\ldots \dot\beta_M}\,,
\end{array}
\end{equation}
where $\partial_{\alpha \dot\beta}=(\sigma^\mu)_{\alpha \dot\beta}\partial_\mu\,$,
$\partial^{\dot\beta\alpha }=(\tilde{\sigma}^\mu)^{\dot\beta\alpha }\partial_\mu\,$
and $\partial_{\beta\dot\gamma}\partial^{\dot\gamma\alpha }= \delta_\beta^\alpha \,\Box\,$,
$\partial^{\dot\alpha\gamma }\partial_{\gamma \dot\beta}= \delta^{\dot\alpha}_{\dot\beta} \,\Box\,$.
Notice that, for a given $L$, the multispinor field $\phi_{(N,M)}$ contains the sequence of spins
$\left(\frac{N+M}{2}=\frac{L}{2},\ldots,\frac{|N-M|}{2}\right)$, as it follows by
looking at the $SU(2)$ representation contents of the finite irreducible representations
of $SL(2,\mathbb{C})$ (see {\it e.g.} \cite{Cor:53}). The nonmaximal $(s<\frac{L}{2})$ spins
are eliminated subjecting  $\phi_{(N,M)}$ to the generalized Lorenz conditions
\begin{equation}
\label{Lor-cond}
\partial^{\dot\beta \alpha} \phi_{\alpha_1\ldots
\alpha_{N-1}\alpha\dot\beta}{}^{\dot\beta_1\ldots \dot\beta_{M-1}}=0\,,
\end{equation}
which follow as well from the formulae \eqref{wv-st}, plus all the tracelessness conditions
which are also consequences of \eqref{wv-st}.

Dirac-Fierz-Pauli equations for spin $s$ can be written in Weyl spinor notation
as equations relating the $\phi_{(L,0)}$ and $\phi_{(L-1,1)}$ multispinor fields
\begin{equation}
\label{DFP-eq1}
\begin{array}{rcl}
i\partial_{\alpha \dot\beta}\phi_{\alpha_1\ldots \alpha_{L-1}}{}^{\dot\beta}&=&
m\,\phi_{\alpha \alpha_1\ldots \alpha_{L-1}} \ ,\\ [7pt]
i\partial^{\dot\beta \alpha}\phi_{\alpha \alpha_1\ldots \alpha_{L-1}}&=&
m\,\phi_{\alpha_1\ldots \alpha_{L-1}}{}^{\dot\beta}\ ;
\end{array}
\end{equation}
alternatively, we can choose eq.~\eqref{DFP-eq} for the multispinors $\phi_{(0,L)}$
and $\phi_{(1.L-1)}$.
The second equation in \eqref{DFP-eq1} can be considered (for $m\neq 0$) as defining the
fields $\phi_{(L-1,1)}$, so the whole set of fields \eqref{wv-st} can be
obtained from the fields $\phi_{(L,0)}$, which
satisfy the massive Klein-Gordon equation and describe spins $s=L/2$
\cite{Cor:53,Wein}.
Indeed, using relations \eqref{DFP-eq} subsequently for the fields
$\phi_{(L-1,1)},...,\phi_{(L-M,M)}$ it may be
shown that all these fields can be expressed in terms of $\phi_{(L,0)}$ by
\begin{equation}
\label{DFP-eq2}
\begin{array}{rcl}
\phi_{\alpha_1\ldots \alpha_{L-M}}{}^{\dot\beta_1\ldots \dot\beta_M}&=&
\frac{i}{m}\,\partial^{\dot\beta_M \alpha_{L-M+1} }
\phi_{\alpha_1\ldots \alpha_{L-M+1}}{}^{\dot\beta_1\ldots \dot\beta_{M-1}}\,=\ldots=
\\ [7pt]
&=&
(\frac{i}{m})^{M}\,\partial^{(\dot\beta_{M}\alpha_{L-M+1}}\partial^{{\dot\beta}_{M-1}\alpha_{L-M+2}}\ldots
\partial^{\dot\beta_{1})\alpha_L}
\phi_{\alpha_1\ldots \alpha_{L}}\,.
\end{array}
\end{equation}
It follows therefore that all the field equations for $\phi_{(N,M)}$
($N$+$M$=$L$) in \eqref{DFP-eq} can be obtained from
an independent pair of linear HS field equations for $\phi_{(N,M)}=\phi_{(L-1,1)}$
and $\phi_{(N,M)}=\phi_{(0,L)}$ or from $\phi_{(L,0)}$ and $\phi_{(1,L-1)}$. In particular if we choose $L$=1
in \eqref{DFP-eq1}, we obtain the standard Dirac equation for a Dirac field
in the Weyl realization as the sum of an undotted spinor and a dotted one,
$\phi_{(1,0)}\oplus\phi_{(0,1)}$ in the our notation.

If $L{=}2$ we obtain the Proca equations
expressed in terms of $\phi_{(1,1)}$, $\phi_{(2,0)}$ and $\phi_{(0,2)}$.
Consider first \eqref{DFP-eq} for $N$=0, $M=2$,
\begin{equation}
\label{Pr-eq1}
i\,\partial_{\alpha_1 \dot\beta_1}\phi^{\dot\beta_1\dot\beta_2} =m\, \phi_{\alpha_1}{}^{\dot\beta_{2}}\,,\qquad
i\,\partial^{\dot\beta_1 \alpha }\phi_{\alpha}{}^{\dot\beta_{2}}=m\, \phi^{\dot\beta_{1}\dot\beta_{2}}.
\end{equation}
Eliminating $\phi^{\dot\beta_1 \dot\beta_2}$, we see that
the vector field $\phi_{\alpha_1 \dot\beta_{2}}$
satisfies the massive Klein-Gordon equation. Further, using the symmetry
in the $\beta_1 \beta_2$ indices, it follows that the two equations
above imply the Lorenz condition (see \eqref{Lor-cond})
\begin{equation}
\label{Lor-cond1}
\partial^{\dot\beta\alpha}\phi_{\alpha\dot\beta}=0\,,
\end{equation}
which eliminates the spin zero part of $\phi_{\alpha \dot\beta}$. Thus, by virtue of
eqs.~\eqref{Pr-eq1}, $\phi_{\alpha \dot\beta}$ is the spin one Proca field $\phi_\mu$ satisfying
$(\Box + m^2)\phi^\mu=0\,,\, \partial_\mu \phi^\mu=0$ (see eq.~\eqref{v-s-4}).
Similarly, if we now consider the case $N$=1=$M$ in eqs.~\eqref{DFP-eq}, we obtain
\begin{equation}
\label{Pr-eq2}
i\,\partial_{\alpha_1 \dot\beta} \phi_{\alpha_2}{}^{\dot\beta}= m\, \phi_{\alpha_1 \alpha_2}\, , \qquad
i\,\partial^{\dot\beta \alpha_1}\phi_{\alpha_1 \alpha_2}  =m \, \phi_{\alpha_2}{}^{\dot\beta} \, .
\end{equation}
As before, the Klein-Gordon equation and the Lorenz condition for the four-vector field
$\phi_{\alpha \dot\beta}$ are contained in  eqs.~\eqref{Pr-eq2}, which again reproduce the
equations satisfied by a Proca field.

   We note that to obtain the Proca equations as a massive extension
of the Maxwell equations it is sufficient to describe the free field dynamics in
terms of the field strength $\phi_{\mu\nu}=\partial_\mu \phi_\nu-\partial_\nu \phi_\mu$.
The tensor $\phi_{\mu\nu}$ may be expressed in terms of its dual and antiselfdual parts,
$\phi_{\mu\nu}\sim (\phi_{(2,0)}$, $\phi_{(0,2)})$.
Using these two bispinor fields one obtains the Proca equations,
$\partial^\mu \phi_{\mu\nu} +m^2\phi_\nu=0$.

\section{Outlook}
\label{beyond}
\setcounter{equation}{0}

We have presented in this paper new {\it massive} particle models in $D{=}3$ and $D{=}4$
spacetimes enlarged in $D{=}3$ by two
($y^{\alpha\beta}_r, r=1,2$) or in $D{=}4$ by three  ($y^{\alpha\dot\beta}_r, r=1,2,3$)
additional copies of Minkowskian four-vector variables and their momenta.
After quantization, the wave functions are defined on $SL(2;\mathbb{K})$ manifolds
($\mathbb{K}=\mathbb{R}$ for $D{=}3$, $\mathbb{K}=\mathbb{C}$ for
$D{=}4$) and describe towers of free massive  HS fields. A natural
extension of these models is the $D{=}6$ case,
in which the wave functions would be defined on the $SL(2;\mathbb{H})$ manifold,
with $12$ real parameters. In such a case, the complex $D{=}4$ twistors
in  Sec.\,3 should be replaced by quaternionic $D{=}6$ twistors
(see {\it e.g.} \cite{LN91}), defined as fundamental spinorial realization of the
$D{=}6$ conformal $SO(6,2)$ group with spinorial quaternionic covering
$U_\alpha(4;\mathbb{H})\simeq O^\ast(8;\mathbb{C})$ group (see {\it e.g.} \cite{Tits,HLM83}).

We would like to point out that it is possible to relate the $D{=}3$ and $D{=}4$ massive models
with $D{=}4$ and $D{=}5$ massless ones by observing that massless fields in $D+1$ spacetimes
become massive in one less dimension $D$ after dimensional reduction and
interpreting the $(D{+}1)$-th momentum component as the mass
in $D$-dimensional spacetime.  There is a link between the description of
helicity in massless theories and spin in massive case; {\it i.e.}
\begin{equation}
\label{4.1}
(D+1)\ \ \textrm{`helicity'}\  \qquad \longrightarrow \qquad D\ \ \textrm{`spin'} \; .
\end{equation}
In particular, the Abelian helicity operator in $D=4$ (see \eqref{1.8}) corresponds
to the  spin operator \eqref{2.7} in $D=3$ and further the $SU(2)$
spin algebra in $D=4$ could be used analogously to describe the
generalized helicity states in $D=5$.
We add that recently it has been pointed out that the symplectic two-form
describing the spin contribution to the  $D=3$ free massive spinning particle
dynamics \cite{MRT13} can be identified with the symplectic two-form describing the helicity
part of particle dynamics for massless $D=4$ particles with nonvanishing helicity
\cite{DuHor}.

One can extend our considerations to the supersymmetric
case. In it, as in the first example of a spinorial
particle model in tensorial spacetime extended
to superspace \cite{BLS00}, the additional variables can be associated
with the so-called tensorial central charges of the supersymmetry
algebras ({\it i.e.}, central but for the Lorentz subalgebra).
These charges play an important role in the theory of
supersymmetric extended objects \cite{AITG:89};
the corresponding central tensorial generators of the superlagebras
act as differential operators on the additional coordinates of the associated
extended superspaces\footnote{For a discussion
of the role of additional coordinates of extended superspaces
see \cite{A-I:99-08} and references therein.}.
From this perspective, the two-twistor models
introduced here can be related with the tensorial central charges
of $N=2$ supersymmetry and the variables of the suitably extended
superspaces. The most general $D=3$ $N=2$
superalgebra extended by tensorial central charges is as follows
\begin{equation}
\label{4.2}
    \{ Q_\alpha^i, Q_\beta^j \} = \delta^{ij} P_{\alpha\beta} +
    (\sigma_1)^{ij} Z^{(1)}_{\alpha\beta} + (\sigma_3)^{ij}
    Z^{(2)}_{\alpha\beta}+ \epsilon^{ij}\epsilon_{\alpha\beta}\tilde Z\,.
\end{equation}
The real vectorial `central' charges $Z^{(1)}_{\alpha\beta}$,
$Z^{(2)}_{\alpha\beta}$ may be considered as the momenta $p^{r}_{\alpha\beta}$
generating the translations of our additional coordinates $y^{\alpha\beta}_r$
($r=1,2$; see \eqref{2.8} and \eqref{2.12a}). In this view, the
first formula \eqref{2.13} takes the form
\begin{equation}
\label{4.3}
Z_{\alpha \beta}^{(r)} = -i
\frac{\partial}{\partial y^{\alpha \beta}_{r}} \ .
\end{equation}

In the $D$=4, $N=2$ supersymmetry algebra with tensorial central charges,
the generators associated with the coordinates listed in \eqref{1.9} (see also Sec.\,3) appear
as part of those of the extended superalgebra
\begin{eqnarray}
\label{4.4a}
\{ Q_\alpha^i, \bar{Q}_{\dot{\beta}}^j \} &=& (\sigma_a)^{ij}
P_{\alpha \dot{\beta}}{}^a \ , \\
\{ Q_\alpha^i, Q_\beta^j \} &=& \delta^{ij} \tilde{Z}_{\alpha\beta}
+(\sigma_1)^{ij} \tilde{Z}_{\alpha\beta}^{(1)} + (\sigma_3)^{ij}
\tilde{Z}_{\alpha\beta}^{(2)}+ \epsilon^{ij}
\epsilon_{\alpha\beta} \tilde{Z}\, \label{4.4b}
\end{eqnarray}
(similarly for ($\{\bar{Q}_{\dot{\alpha}}^i, \bar{Q}_{\dot{\beta}}^j \}$),
where the $16$ generators $P_{\alpha \dot{\beta}}^a$ are real and
the $10$ generators $\tilde{Z}_{\alpha\beta}$,
$\tilde{Z}_{\alpha\beta}^{(1)}$, $\tilde{Z}_{\alpha\beta}^{(2)}$,
$\tilde{Z}$ are complex ({\it i.e.} there are $36$ bosonic real
generators).
In our $D=4$ model we have only used the
sixteen coordinates $y_{\alpha\dot\beta}^a$ (see eq.~\eqref{2.8})
associated with $P^a_{\alpha \dot{\beta}}$, $a$=0,1,2,3
and the remaining $10$ complex tensorial charges were put
equal to zero. Let us observe that for $N=1$ only the first term in
the {\it r.h.s.} of the relation \eqref{4.4b} survives and describes the tensorial central
charges used in \cite{BL99,BLS00,V+,PST03}. If $N=2$ we also note that the generator $\tilde{Z}$
in \eqref{4.2} and \eqref{4.4b} that we did not include in our considerations
is a truly central one (it is a Lorentz scalar).
This generator, associated with a scalar central coordinate,
has eigenvalues characterizing the mass; its role
in $N=2$ massive superparticle model was elucidated
long ago \cite{dAL82}.

The models discussed in this paper give the same mass for all HS fields,
which of course is very restrictive.
In a physical HS case, when considering
$e.g.$ spin excitations in string theory, the masses are spin-dependent.
They lie on a Regge trajectory, which in the general case can be described
by replacing the constant mass by an spin-dependent function
$m=m(s)$ (usually linear). In this case, the constant $m$ in
the mass-shell condition should be replaced by a spin-dependent
operator (see \eqref{3.12}, \eqref{3.13}, \eqref{3.48}) {\it i.e.},
\begin{equation}
\label{4.5}
m^2\qquad \rightarrow \qquad m^2 (\vec{S}^2) \ .
\end{equation}
In the twistor formulation, the spinorial mass shell conditions
(eq.~\eqref{1.10} in $D=4$) may be considered as `complex roots' of the
standard mass shell condition. It is an interesting problem to
see how to introduce, in the complex mass parameter $M$ appearing
in eqs.~\eqref{1.10}, a dependence on the twistor variables that could lead
to HS multiplets with masses on a Regge trajectory.

Another problem worth studying is the description of {\it interacting}
{\it massive} HS theories. To this aim, one could follow the Fradkin-Vasiliev prescription
for massless HS fields \cite{FradVas87} and introduce in our formalism the AdS
radius {\it i.e.}, one could generalize the set of coordinates
($x_\mu$, $y^r_\mu$) (see \eqref{1.9}) to the case where $x_\mu$ is endowed with
a  constant spacetime curvature. However, in the interacting massive HS
theory, the finite $AdS$
radius is not necessary because it is possible to rescale
the derivatives in higher order terms by using the mass parameter.
Indeed, the cosmological constant $\Lambda$ and mass parameter $m$
play an analogous role in the field equations, as reflected {\it e.g.} in the shift
$m^2\rightarrow m^2\pm \frac32\Lambda^2$
(see e.g. \cite{PonVas2010}) which appears when the KG equation is formulated in $(A)dS$
spacetime\footnote{The problem with the limit $m\to 0$ for flat ($\Lambda=0$) HS fields is
also reflected in the appearance of Van\,Dam-Veltman-Zakharov discontinuity \cite{DaVel70,Zakh70}
in massive gravity models. However, if $\Lambda\neq 0$, in the limit $m\to 0$
such a discontinuity does not exist \cite{Por01,DesWal01}. }.
Theefore, one can conclude that massive interacting
HS theories should already exist on a flat Minkowski background.

The interacting massless HS theory is usually described as a HS theory,
also called HS gravity\footnote{
We recall that gravity can be described as gauge theory of massless $s=2$ fields,
with local gauge transformations realized as spacetime diffeomorphism
(see e.g. \cite{OgiPolu65}).
},
with nonlinearities generated {\it e.g.} by
non-Abelian HS field strengths. In such framework the mass can be introduced in two ways.

{\bf 1)}
In the first one, the mass parameter appears as a consequence of the spontaneous symmetry breaking
of HS local gauge symmetries, due to the coupling of the HS fields to supplementary
Stueckelberg fields carrying spontaneously broken local gauge degrees of
freedom \cite{Stuec38,Luk65,Ru04,Met-Stuec,BeiMorS04,BiaHes05}. In the Vasiliev
formalism (see e.g. \cite{Vas2012,DidSk14,Vasil14}), the HS gauge connection fields
are given by a vectorial master field $\omega_\mu(x;y^{\alpha}_{i},\bar y^{\dot\alpha i})$;
a scalar master field $C(x;y^{\alpha}_{i},\bar y^{\dot\alpha i})$ encodes,
besides the gauge-invariant HS curvatures, new degrees of freedom that
describe the low spin ($s=0$ and $s=\frac12$) matter.
The introduction of a  Stueckelberg-Higgs mechanism requires
new local symmetries described by a new pure gauge scalar master field
$\widetilde C(x;y^{\alpha}_{i},\bar y^{\dot\alpha i})$. This field accounts
for the set of spontaneously broken local gauges given as covariantized shifts of
the spacetime fields defined by the spinorial Taylor expansion of
$\widetilde C(x;y^{\alpha}_{i},\bar y^{\dot\alpha i})$.
The masses for the the various spins are obtained through
the particular gauge fixings that replace the spacetime dependent
field components of $\widetilde C(x;y^{\alpha}_{i},\bar y^{\dot\alpha i})$ by parameters.

  The idea of obtaining nonvanishing mass via Stueckelberg-Higgs mechanism in massless
HS gauge theory was also proposed in (super)string theory to generate
an infinite collection of massive states which lie on Regge trajectories.
Recently a generalization of HS algebras has been proposed under
the name of multiparticle extension of HS symmetries \cite{Vasil13,GelVas13}~\footnote{
The standard HS algebras are described by the linear basis of various enveloping Heisenberg
algebras, with canonical generators represented by quantized vectorial
or spinorial (twistorial) coordinates (see e.g. \cite{Vasil88,KonsVas90,Boul13}).
The multiparticle extension of HS algebra is described as ``doubly infinite''
enveloping standard HS algebras.}.
This permits to look at a string as an infinite collection of interacting HS multiplets
which, in a miraculous way, seem to provide a finite
(or at least renormalizable) example of interacting massive HS field theory.

{\bf 2)}
In the second case,  the mass parameter in the massive HS theory produces a
``hard'' breaking of local HS gauge symmetries and
the massive HS fields may be considered as describimg the sector of non-gauge HS matter,
entering in HS gravity equations at r.h.s. of HS-extended Einstein equations.
These HS fields would be related with HS massive currents, which
for massless HS fields are described by current master fields of rank two
with double number of spinorial coordinates \cite{GelVas2005,GelVas2010,Vas2012,GelVas13}.
This last property is also a feature of our massive master HS fields,
with double number of twistorial coordinates. This suggests that the massive currents
could be introduced in a natural way by using a master field analogue of the old
current-field identities idea \cite{LeeWeiZum67,Saku69}.

The current master fields describe the ``conformal side'' in the realization
of HS/CFT duality in Vasiliev theory \cite{GelVas2005,GelSVas2008,GelVas2010,Vas2012,GelVas13}.
As shown by Flato and Fronsdal \cite{FlatoFron78},
the binary products of $D{=}3$ singletons describe the $D{=}4\rightarrow D{=}3$
reduction of the $D{=}4$ multiplet of free AdS HS fields; analogously, in HS field theory
the bilinear products of $D{=}3$ massless master fields describe the conserved HS currents
that  can be identified with the $D{=}3$ holomorphic
boundary of the $D{=}4$ massless AdS master field. For massive HS fields
the conformal HS symmetry is broken, which leads to the nonconservation of the HS conformal
currents and the deformation of HS AdS/CFT duality picture that is obtained
for massless HS theory. The way in which the nonvanishing mass of HS fields modifies
the known massless HS AdS/CFT duality scheme will be a subject for our
future research.

In Sects.\,2 and 3 we considered wave functions depending on the spinors
$\lambda_\alpha^i$ in $D{=}3$ and $\pi_\alpha^i$ in $D{=}4$. These spinorial momentum coordinates,
due to the constraints  \eqref{2.4} and \eqref{1.10}, describe $D{=}3$ and $D{=}4$
Lorentz group manifolds. Considering the differential realization of the spinorial variables,
$\hat\lambda_\alpha^i=-i\partial/\partial y^\alpha_i$ in $D{=}3$ and
$\hat\pi_\alpha^i=-i\partial/\partial y^\alpha_i$ in $D{=}4$, we can
compare our results with the unfolding equations for free massless fields \cite{BLS00,V+,PST03}.
In the massless case, following \cite{Frons}, one considers a tensorial extension of spacetime
$x^\mu\rightarrow (x^\mu,y^{[\mu\nu]})$ that permits the introduction of $Sp(8)$-covariant fields
in Minkowski space with all $D{=}4$ helicities belonging to an irreducible $Sp(8)$ representation.
In our massive model, the extension $x^\mu\rightarrow (x^\mu,y_r^{\mu})$
of spacetime is linked to the introduction of Lorentz-covariant $SU(2)$ frames
(see e.g. \eqref{xxx}, \eqref{3.7}) describing all $D{=}4$ spin degrees of freedom\footnote{
The introduction of Lorentz frames to describe spin kinematics goes back to Souriau
\cite{Sour,Kunz}. Unfortunately we were not able to introduce the manifold $(x^\mu,y_r^{\mu})$
in a group-theoretical way, as {\it e.g.} $(x^\mu,y^{[\mu\nu]})$
can be defined as a parabolic coset of $Sp(8)$ \cite{V+,PST03}. }.
However, both extensions of the $D{=}4$ Minkowski spacetime described above
are rather introduced by symmetry arguments and, in both cases,
these extensions are not mandatory to define the free dynamics of HS master fields.
If we eliminate the auxiliary spacetime variables by solving the auxiliary
unfolding equations (\eqref{3.23} with $a=1,2,3$ in massive case), we are left with
a truncated form of the unfolding equation with only standard spacetime derivatives.
In the massive case the free HS master fields are described by the
unfolded equation \eqref{3.23} for $a=0$ ($y_0{}^{{\dot \beta} \alpha }$
are the spacetime coordinates)
supplemented with the mass quantum constraints \eqref{1.10}:
\begin{equation}
\label{unf-app}
\left( i \,\partial_{\alpha\dot{\beta}}
-\frac{\partial^2}{\partial y^\alpha_i \partial \bar y^{\dot\beta i}}
\right)\Psi(x, y,\bar{y}) = 0 \; ,
\end{equation}
\begin{equation}
\label{mass-app}
\left( \frac{\partial^2}{\partial y^\alpha_i \partial y_{\alpha}^{i}}
- 2M
\right)\Psi(x, y,\bar{y}) = 0 \; ,\qquad
\left( \frac{\partial^2}{\partial \bar y^{\dot\alpha}_{i} \partial \bar y_{\dot\alpha}^{i}}
- 2\bar M
\right)\Psi(x, y,\bar{y}) = 0\, .
\end{equation}
Equations \eqref{unf-app} and \eqref{mass-app} describe free massive HS fields in flat
Minkowski space; by a suitable (A)dS covariantization of the vectorial and spinorial derivatives
in \eqref{unf-app}, \eqref{mass-app} the master field equations describing
free massive HS fields in (A)dS spacetime may be obtained. Further,
one can study the couplings of massive HS fields with full HS gravity background
as well as dynamical gauge fields.


Finally,  we note that the use of a generalized spacetime with vector
variables going beyond the standard spacetime vector is also an important ingredient of the
BRST approach to the Lagrangian formulation of HS fields\footnote{SF thanks I.L.\,Buchbinder for
a clarifying discussion
about this approach. BRST techniques are not considered in this paper.} developed in
\cite{PTs98,BPTs01,BuchPT01,BuchKT15}.

\subsection*{Acknowledgments}
The authors would like to thank M.A.\,Vasiliev for valuable remarks.
This paper has been partially supported by research grants from the Spanish
MINECO (CONSOLIDER CPAN-CSD2007-00042) and by the Polish National
Science Center project 2013/09/B/ST2/02205, 2014/13/B/ST2/04043 and
Maestro grant 2013/10/A/ST2/00106.
S.F. acknowledges support from the RFBR grants 12-02-00517, 13-02-90430 and
a grant of the Bogoliubov-Infeld Programme.

\renewcommand\theequation{A.\arabic{equation}} \setcounter{equation}0
\section*{Appendix A: Notation}

\subsection*{A1. $D$=3 spacetime}

The spacetime metric is $\eta_{\mu\nu}={\rm diag}(+1,-1,-1)$.
Dirac spinor indices are labeled by $\alpha=1,2$ and we use mostly $D$=3 real
Majorana spinors. In particular, the twistor variables
$({\lambda^i_\alpha})^\ast=\lambda^i_\alpha$, $i=1,2$
are real.

We use the following real Majorana realization for the $\gamma$-matrices:
\begin{equation}
\label{3gamma-Cl}
\{\gamma^\mu,\gamma^\nu\}=-2\,\eta^{\mu\nu}\;,
\end{equation}
\begin{equation}
\label{3gamma}
(\gamma_\mu)_\alpha{}^\beta\,:\qquad \gamma_0=
i\sigma_2\,,\quad \gamma_1=\sigma_1\,,\quad\gamma_2=\sigma_3\,,
\end{equation}
where $\sigma_1,\sigma_2,\sigma_3$ are usual Pauli matrices.
In this realization the antisymmetric the charge conjugation matrix
$C_{\alpha\beta}=\epsilon_{\alpha\beta}$ coincides with the
matrix $\gamma^0$. Thus, spinor indices are raised and lowered by
$\lambda^{\alpha i}=\epsilon^{\alpha\beta}\lambda^i_\beta$,
$\lambda^{ i}_\alpha=\epsilon_{\alpha\beta}\lambda^{\beta i}$,
where $\epsilon_{12}=\epsilon^{21}=1$.
The matrices
\begin{equation}
\label{3gamma-1}
(\gamma_\mu)_{\alpha\beta}=
\epsilon_{\beta\gamma}(\gamma_\mu)_\alpha{}^\gamma\,:\qquad (\gamma_0)_{\alpha \beta}={\bf 1}_2\,,\quad
(\gamma_1)_{\alpha \beta}=\sigma_3\,,\quad(\gamma_2)_{\alpha \beta}=\sigma_1
\end{equation}
form a basis for the $2\times 2$ symmetric matrices. In particular,
\begin{equation}
\label{3gamma-1a}
(\gamma^\mu)_{\alpha\beta}(\gamma_\nu)^{\alpha\beta}=2\,\delta^\mu_\nu\,.
\end{equation}
As a result ($A_\mu$ and $B_\mu$ are three-vectors)
\begin{equation}
\label{v-s-3}
A_{(\alpha\beta)}={\textstyle\frac{1}{\sqrt{2}}}\,A_\mu(\gamma^\mu)_{\alpha\beta}\,,
\qquad A_\mu ={\textstyle\frac{1}{\sqrt{2}}}\,A_{\alpha\beta}(\gamma_\mu)^{\alpha\beta} \,,
\end{equation}
and
\begin{equation}\label{v-s-3a}
A^\mu B_\mu =A^{\alpha\beta}B_{\alpha\beta} \,.
\end{equation}

In Sec.~\ref{desc} we also use a complex representation of $D=3$ Dirac-Clifford algebra,
which is obtained from the Majorana realization \eqref{3gamma} by the similarity
transformation \eqref{sim-tr}: $\gamma_\mu\,\to\,U\gamma_\mu U^{-1}$.
In such a realization of $D=3$ Dirac algebra
we use $SU(1,1)$ as the $Spin(2,1)$ group, and the $\gamma$-matrices take the form
\begin{equation}
\label{3gamma-c}
(\gamma_\mu)_\alpha{}^\beta\,:\qquad \gamma_0=
i\sigma_3\,,\quad \gamma_1=\sigma_1\,,\quad\gamma_2=-\sigma_2\,,
\end{equation}
\begin{equation}
\label{3gamma-1-c}
(\gamma_\mu)_{\alpha\beta}=
\epsilon_{\beta\gamma}(\gamma_\mu)_\alpha{}^\gamma\,:\qquad (\gamma_0)_{\alpha \beta}=
 -i\sigma_1\,,\quad
(\gamma_1)_{\alpha \beta}=
\sigma_3\,,\quad(\gamma_2)_{\alpha \beta}=i{\bf 1}_2 \,.
\end{equation}

We will proceed similarly for matrices with internal $i,j$ indices. In particular,
$\lambda_{\alpha}^{i}=\epsilon^{ij}\lambda_{\alpha j}$,
$\lambda_{\alpha i}=\epsilon_{ij}\lambda_{\alpha}^{j}$,
where $\epsilon_{ij}$ and $\epsilon^{ij}$ are defined by $\epsilon_{12}=\epsilon^{21}=1$.
Also, we will use the matrices ($a=0,1,2$)
acting of internal indices
\begin{eqnarray}
\label{3gamma-2}
(\gamma_a)_i{}^j\,:&&\qquad \gamma_0=i\sigma_2\,,\quad \gamma_1=\sigma_1\,,\quad\gamma_2=\sigma_3\,,\\
\label{3gamma-3}
(\gamma_a)_{ij}=\epsilon_{jk}(\gamma_a)_i{}^k\,:&&\qquad \gamma_0={\bf 1}_2\,,\quad \gamma_1=\sigma_3\,,\quad\gamma_2=-\sigma_1
\end{eqnarray}

\subsection*{A2. $D$=4 spacetime}

The spacetime metric is $\eta_{\mu\nu}={\rm diag}(+1,-1,-1,-1)$.
We shall use the two-component Weyl spinor notation.\~In particular,
four-vector quantities are defined in terms of spinors as $x_{\alpha\dot{\beta}} =
x_\mu\sigma^\mu_{\alpha\dot\beta}$, where
\begin{equation}
\label{sigma-1}
(\sigma_\mu)_{\alpha\dot\beta}=({\bf 1_2};\sigma_1,\sigma_2,\sigma_3)_{\alpha\dot\beta}
\end{equation}
and $\sigma_1,\sigma_2,\sigma_3$ are the Pauli matrices. Spinor indices are raised and
lowered by $\epsilon_{\alpha\beta}$, $\epsilon^{\alpha\beta}$, $\epsilon_{\dot\alpha\dot\beta}$,
$\epsilon^{\dot\alpha\dot\beta}$ with nonvanishing components
$\epsilon_{12}=-\epsilon_{21}=\epsilon^{21}=-\epsilon^{12}=1$. As the result, the matrices
\begin{equation}
\label{sigma-2}
(\tilde\sigma_{\mu})^{\dot\alpha\beta}=\epsilon^{\dot\alpha\dot\delta}\epsilon^{\beta\gamma}
(\sigma_\mu)_{\gamma\dot\delta}=({\bf 1_2};-\sigma_1,-\sigma_2,-\sigma_3)^{\dot\alpha\beta} \; ,
\end{equation}
satisfy
\begin{equation}
\label{sigma-3}
\sigma^\mu_{\alpha\dot\gamma}\tilde\sigma^{\nu\,\dot\gamma\beta}+
\sigma^\mu_{\alpha\dot\gamma}\tilde\sigma^{\nu\,\dot\gamma\beta}=
2\,\eta^{\mu\nu}\delta^\beta_\alpha \quad,\quad
\sigma^\mu_{\alpha\dot\beta}\tilde\sigma_\nu^{\dot\beta\alpha}=2\,\delta^\mu_\nu\,.
\end{equation}
The Dirac matrices are given by
\begin{equation}
\label{Dir-matr}
\gamma_\mu=\left(
             \begin{array}{cc}
               0 & \sigma_\mu \\
               -\tilde\sigma_{\mu} & 0 \\
             \end{array}
           \right),
\qquad \{\gamma^\mu,\gamma^\nu\}=- 2\,\eta^{\mu\nu}
\,.
\end{equation}
The link between Minkowski four-vectors and spinorial quantities is given by
\begin{equation}
\label{v-s-4}
A_{\alpha\dot\beta}={\textstyle\frac{1}{\sqrt{2}}}\,A_\mu(\sigma^\mu)_{\alpha\dot\beta}\,,
\qquad A_\mu ={\textstyle\frac{1}{\sqrt{2}}}\,A_{\alpha\dot\beta}(\tilde\sigma_\mu)^{\dot\beta\alpha} \,,
\end{equation}
so that
\begin{equation}
\label{v-s-4a}
A^\mu B_\mu =A^{\alpha\dot\beta}B_{\alpha\dot\beta} \,.
\end{equation}

Similar matrices are used in internal space with indices $i$, $j$. At
this point it is necessary to make a comment. There are two methods to indicate
the complex conjugate spinor representation. The first one uses dotted indices
as in \eqref{sigma-1}. The second method, often used for $SU(2)$, raises and lowers
two-spinor indices. We use the second method for matrices in internal space.
So, we use matrices
\begin{equation}
\label{sigma-5}
(\sigma_a)_i{}^j=(\sigma_0;\sigma_r)_i{}^j=({\bf 1_2};\sigma_1,\sigma_2,\sigma_3)_i{}^j\,.
\end{equation}
In these matrices indices are raised and lowered by $\epsilon_{ij}$ and
$\epsilon^{ij}$ with components $\epsilon_{12}=\epsilon^{21}=1$;
under complex conjugation the position of these indices is exchanged
{\it e.g.}, $(\pi_\alpha^i)^\ast=\bar\pi_{\dot\alpha i}$.

\renewcommand\theequation{B.\arabic{equation}} \setcounter{equation}0
\section*{Appendix B: From $D=3$ spinorial to $D=4$ vectorial  particle model}

The action \eqref{2.8} proposed in this paper in the case $c=1$ becomes
$\overline{SO(2,2)} =\overline{SO(2,1)}\otimes \overline{SO(2,1)_{int}} =
SL(2;\mathbb{R})\otimes SL(2;\mathbb{R})_{int}$-invariant, where the
indices $\alpha,\beta$ describe the $D=3$ Lorentz spinor group, and
$i,j$ the `internal' $\overline{SO(2,1)_{int}}$ indices.

Let the $SO(2,1)$ $N$-spinors $\varphi_{\alpha_1\ldots\alpha_N}$,
symmetrical in $\alpha_1\ldots\alpha_N$, be denoted by $(\frac{N}{2},0)$, and
$SO(2,1)_{int}$ $L$-spinors by $\phi^{i_1\ldots i_M}$ by $(0,\frac{L}{2})$.
General $SO(2,2)$ spinors $(\frac{N}{2},\frac{L}{2})=(\frac{N}{2},0){\otimes}(0,\frac{L}{2})$
will then be denoted by $\psi_{\alpha_1\ldots\alpha_N}^{i_1\ldots i_M}$;
spinors $(\frac{N}{2},\frac{N}{2})$ are then equivalent to
$SO(2,2)$ $N$-tensors. In particular the basic $D=3$, $N=2$  spinors $\lambda_\alpha^i$
in our model describe the $SO(2,2)$ vector
\begin{equation}
\label{22-vec}
\lambda_A \cong (\sigma_A)^\alpha{}_i \lambda_\alpha^i\ ,
\end{equation}
where $A$ denotes the $SO(2,2)$ four-vector indices
and $(\sigma_A)^\alpha{}_i$ are the $SO(2,2)$ $\sigma$-matrices
analogous to the $SO(3,1)$ matrices in \eqref{sigma-1}. The
extended spacetime coordinates $y_a^{\alpha\beta}$ (see \eqref{2.8}) and
the variables  $u^a_{\alpha\beta}$ (see \eqref{2.3}) describe
second order $SO(2,2)$ tensors $(1,1)$.
Further one can show
that first three terms in \eqref{2.8} describe the $SO(2,2)$ invariant
contraction of two $SO(2,2)$ tensors $(1,1)$, and the fourth term
is the contraction of two $SO(2,2)$ four-vectors. The mass-shell condition
is defined by the $SO(2,2)$-invariant scalar length of the
$SO(2,2)$ four-vector $\lambda_A$ (see \eqref{22-vec}).

Our model \eqref{2.8} defines therefore the extension of the $N=2$ $D=3$ Shirafuji
model to the vectorial model in $SO(2,2)$ tensorial space $y_a^{\alpha\beta}$.
Such a model cannot be however extended to a corresponding $O(3,3)$ twistorial model,
because $SO(3,3)$ twistors are described by the pair of primary
$SO(2,2)$ spinors $(\frac{1}{2},0){\oplus}(0,\frac{1}{2})$ which we denote as
$(\lambda_\alpha,\lambda^i)$ and $(\omega_\alpha,\omega^i)$.
The second pair of spinors should be defined in terms of $SO(2,2)$ spacetime coordinates
$x_\alpha^i$ by the $SO(2,2)$ incidence relations
\begin{equation}
\label{22-inc}
\omega_\alpha =x_\alpha^i \lambda_i\ ,\qquad \omega^i =x_\alpha^i \lambda^\alpha\,.
\end{equation}
However, in this paper we did not use neither of the simple spinors $\lambda_\alpha$, $\lambda^i$
nor the incidence relations \eqref{22-inc} {\it i.e.}, if we pass to an $SO(2,2)$ interpretation of
our model \eqref{2.8}, we loose the corresponding $SO(2,2)$ twistorial formulation.


\begin{thebibliography}{20}

\bibitem{Frons}
C.\,Fronsdal, {\it Massless particles, orthosymlectic symmetry and another Type
of Kaluza-Klein theory} in {\it Essays on Supersymmetry}, (Mathematical
Physics Studies, v.8), Reidel, Dordrecht, 1986.

\bibitem{BL99}
I.\,Bandos, J.\,Lukierski,
{\it Tensorial central charges and new superparticle models with fundamental spinor coordinates},
Mod. Phys. Lett. {\bf A14} 1999) 1257 [arXiv:hep-th/9811022].

\bibitem{BLS00} BLS00,V+,PST03
I.\,Bandos, J.\,Lukierski, D.\,Sorokin,
{\it Superparticle models with tensorial central charges},
Phys. Rev. {\bf D61}, 045002 (2000) [arXiv:hep-th/9904109].

\bibitem{V+}
M.A.\,Vasiliev,
{\it Conformal higher spin symmetries of 4d massless supermultiplets and $osp(L,2M)$ invariant equations
in generalized (super)space},
Phys. Rev. {\bf D66} (2002) 066006 [arXiv:hep-th/0106149].

\bibitem{PST03}
M.\,Plyushchay, D.\,Sorokin, M.\,Tsulaia,
{\it Higher spins from tensorial charges and $OSp(N|2n)$ symmetry},
JHEP {\bf 0304} (2003) 013 [arXiv:hep-th/0301067].

\bibitem{FI}
S.\,Fedoruk, E.\,Ivanov,
{\it Master higher-spin particle},
Class. Quant. Grav. {\bf 23} (2006) 5195 [arXiv:hep-th/0604111].

\bibitem{PC72}
R.\,Penrose, M.A.H.\,MacCallum,
{\it Twistor theory: an approach to the quantization of fields and spacetime},
Phys. Rept. {\bf 6} (1972) 241.

\bibitem{S83}
T.\,Shirafuji,
{\it Lagrangian mechanics of massless particles with spin},
Prog. Theor. Phys. {\bf 70} (1983) 18.

\bibitem{Fr78}
C.\,Fronsdal,
{\it Massless fields with integer spin},
Phys. Rev. {\bf D18} (1978) 3624.

\bibitem{Sor04}
D.\,Sorokin,
{\it Introduction to the classical theory of higher spins},
AIP Conf. Proc. {\bf 767} (2005) 172 [arXiv:hep-th/0405069].

\bibitem{Jev11}
A.\,Jevicki, K.\,Jin, Q.\,Ye,
{\it Collective dipole model of AdS/CFT and higher spin gravity},
J. Phys. {\bf A44} (2011) 465402 [arXiv:1106.3983 [hep-th]].

\bibitem{Jev14}
R.\,de Mello\,Koch, A.\,Jevicki, J.P.\,Rodrigues, J.\,Yoon,
{\it Holography as a gauge phenomenon in higher spin duality},
arXiv:1408.1255 [hep-th].

\bibitem{BBCL}
A.K.H.\,Bengtsson, I.\,Bengtsson, M.\,Cederwall, N.\,Linden,
{\it Particles, superparticles and twistors},
Phys. Rev. {\bf D36} (1987) 1766.

\bibitem{BCed}
I.\,Bengtsson, M.\,Cederwall,
{\it Particles, twistors and the division algebras},
Nucl. Phys. {\bf B302} (1988) 81.

\bibitem{Tits}
J.\,Tits,
{\it Tabellen zu den einfachen Lie-Gruppen und ihren Darstellungen},
Lecture Notes in Mathematics, Vol. 40, Springer-Verlag, Berlin, 1967.

\bibitem{Ferber}
A.\,Ferber,
{\it Supertwistors and conformal supersymmetry},
Nucl. Phys. {\bf B132} (1978) 55.

\bibitem{EisSol}
Y.\,Eisenberg,
{\it Supertwistors and Superpoincare invariant actions for all linearized extended
supersymmetric theories in four-dimensions},
Mod. Phys. Lett. {\bf A4} (1989) 195;
Y.\,Eisenberg, S.\,Solomon,
{\it (Super)field theories from (super)twistors},
Phys. Lett. {\bf B220} (1989) 562.

\bibitem{Vas1989}
M.A.\,Vasiliev,
{\it Consistent equations for interacting massless fields of all spins
in the first order in curvatures},
Annals Phys. {\bf 190} (1989) 59.

\bibitem{Vas1991}
M.A.\,Vasiliev,
{\it Algebraic aspects of the higher spin problem},
Phys. Lett. {\bf B257} (1991) 111;
{\it More on equations of motion for interacting massless fields of all spins in (3+1)-dimensions},
Phys. Lett. {\bf B285} (1992) 225.

\bibitem{Vas2001}
M.A.\,Vasiliev,
{\it Relativity, causality, locality, quantization and duality
in the $Sp(2M)$ invariant generalized spacetime},
in {\it Multiple facets of quantization and supersymmetry}, Michael Marinov Memorial Volume,
Eds. M.\,Olshanetsky and A.\,Vainshtein, World Scientific, 2002,
826-872  [arXiv:hep-th/0111119].

\bibitem{GelVas2005}
O.A.\,Gelfond, M.A.\,Vasiliev,
{\it Higher rank conformal fields in the Sp(2M) symmetric generalized space-time},
Theor. Math. Phys. {\bf 145} (2005) 1400 [arXiv:hep-th/0304020].

\bibitem{GelSVas2008}
O.A.\,Gelfond, E.D.\,Skvortsov, M.A.\,Vasiliev,
{\it Higher spin conformal currents in Minkowski space},
Theor. Math. Phys. {\bf 154} (2008) 294 [arXiv:hep-th/0601106].
 	
\bibitem{GelVas2010}
O.A.\,Gelfond, M.A.\,Vasiliev,
{\it Unfolded equations for current interactions of 4d massless fields
as a free system in mixed dimensions},
[arXiv:1012.3143 [hep-th]].

\bibitem{Vas2012}
M.A.\,Vasiliev,
{\it Holography, unfolding and higher-spin theory},
J. Phys. {\bf A46} (2013) 214013 [arXiv:1203.5554 [hep-th]].

\bibitem{Vas1994}
M.A.\,Vasiliev,
{\it Unfolded representation for relativistic equations in (2+1) anti-De Sitter space},
Class. Quant. Grav. {\bf 11} (1994) 649.

\bibitem{BBB1983}
A.K.H.\,Bengtsson, I.\,Bengtsson, L.\,Brink,
{\it Cubic interaction terms for arbitrary spin},
Nucl. Phys. {\bf B227} (1983) 31.

\bibitem{BBD1985}
F.A.\,Berends, G.J.H.\,Burgers, H.\,van\,Dam,
{\it On the theoretical problems in constructing interactions involving
higher spin massless particles},
Nucl. Phys. {\bf B260} (1985) 295.

\bibitem{DidSk14}
V.E.\,Didenko, E.D.\,Skvortsov,
{\it Elements of Vasiliev theory},
[arXiv:1401.2975 [hep-th]].

\bibitem{Vasil14}
M.A.\,Vasiliev,
{\it Higher-Spin theory and space-time metamorphoses},
Lect. Notes Phys. {\bf 892} (2015) 227 [arXiv:1404.1948 [hep-th]].

\bibitem{ProVas1997}
A.V.\,Barabanshchikov, S.F.\,Prokushkin, M.A.\,Vasiliev,
{\it Free equations for massive matter fields in (2+1)-dimensional anti-de Sitter
space from deformed oscillator algebra},
Theor. Math. Phys. {\bf 110} (1997) 295 [arXiv:hep-th/9609034].

\bibitem{ProVas1999}
S.F.\,Prokushkin, M.A.\,Vasiliev,
{\it Higher spin gauge interactions for massive matter fields in 3-D AdS space-time},
Nucl. Phys. {\bf B545} (1999) 385 [arXiv:hep-th/9806236].

\bibitem{ShaVas2000}
O.V.\,Shaynkman, M.A.\,Vasiliev,
{\it Scalar field in any dimension from the higher spin gauge theory perspective},
Theor. Math. Phys. {\bf 123} (2000) 683 [arXiv:hep-th/0003123]. 	

\bibitem{Zin2008}
Yu.M.\,Zinoviev,
{\it Frame-like gauge invariant formulation for massive high spin particles},
Nucl. Phys. {\bf B808} (2009) 185 [arXiv:0808.1778 [hep-th]].
 	
\bibitem{BIS2008}
N.\,Boulanger, C.\,Iazeolla, P.\,Sundell,
{\it Unfolding mixed-symmetry fields in AdS and the BMV
conjecture: I. General formalism},
JHEP {\bf 0907} (2009) 013 [arXiv:0812.3615 [hep-th]].

\bibitem{PonVas2010}
D.S.\,Ponomarev, M.A.\,Vasiliev,
{\it Frame-like action and unfolded formulation for massive
higher-spin Fields},
Nucl. Phys. {\bf B839} (2010) 466 [arXiv:1001.0062 [hep-th]].

\bibitem{BPSS2014}
N.\,Boulanger, D.\,Ponomarev, E.\,Sezgin, P.\,Sundell,
{\it New unfolded higher spin systems in $AdS_3$},
[arXiv:1412.8209 [hep-th]].
 	
\bibitem{FZ03}
S.\,Fedoruk, V.G.\,Zima,
{\it Bitwistor formulation of massive spinning particle},
Journal of Kharkov University {\bf 585} (2003) 39, {\tt arXiv:hep-th/0308154};
{\it Bitwistor formulation of the spinning particle},
the Proceedings of the International Workshop {\it Supersymmetry and Quantum Symmetries},
Dubna, 24-29 July, 2003 [arXiv:hep-th/0401064].

\bibitem{FFLM06}
S.\,Fedoruk, A.\,Frydryszak, J.\,Lukierski, C.\,Miquel-Espanya,
{\it Extension of the Shirafuji model for massive particles with spin},
Int. J. Mod. Phys. {\bf A21} (2006) 4137 [arXiv:hep-th/0510266].

\bibitem{AIL08}
J.A.\,de\,Azc\'arraga, J.M.\,Izquierdo, J.\,Lukierski,
{\it Supertwistors, massive superparticles and $\kappa$-symmetry},
JHEP {\bf 0901} (2009) 041 [arXiv:0808.2155 [hep-th]].

\bibitem{MRT13}
L.\,Mezincescu, A.J.\,Routh, P.K.\,Townsend,
{\it Supertwistors and massive particles},
Annals Phys. {\bf 346} (2014) 66
[arXiv:1312.2768 [hep-th]].

\bibitem{T77}
K.P.\,Tod,
{\it Some symplectic forms arising in twistor theory},
Rept. Math. Phys. {\bf 11} (1977) 339.

\bibitem{P79}
Z.\,Perj\'{e}s,
{\it Twistor variables of relativistic mechanics},
Phys. Rev.  {\bf D11} (1975) 2031;
{\it Unitary space of particle internal states},
Phys. Rev.  {\bf D 20} (1979) 1857.

\bibitem{H79}
L.P.\,Hughston, {\it Twistors and particles},
Lecture Notes In Physics, Vol. {\bf 97}, Springer-Verlag, Berlin, 1979.

\bibitem{B84}
A.\,Bette, {\it On a point-like relativistic and spinning particle},
J. Math. Phys. {\bf 25} (1984) 2456;
{\it Directly interacting massless particles: A twistor approach},
J. Math. Phys. {\bf 37} (1996) 1724
[arXiv:hep-th/9601017].

\bibitem{BALM04}
A.\,Bette, J.A.\,de\,Azc\'arraga, J.\,Lukierski, C.\,Miquel-Espanya,
{\it Massive relativistic particle model with spin and electric charge from two twistor dynamics},
Phys. Lett. {\bf B595} (2004) 491
[arXiv:hep-th/0405166].

\bibitem{AFLM06}
J.A.\,de\,Azc\'arraga, A.\,Frydryszak, J.\,Lukierski, C.\,Miquel-Espanya,
{\it Massive relativistic particle model with spin from free two-twistor dynamics
and its quantization},
Phys. Rev. {\bf D73} (2006) 105011 [arXiv:hep-th/0510161].

\bibitem{FL14}
S.\,Fedoruk, J.\,Lukierski,
{\it Massive twistor particle with spin generated by Souriau-Wess-Zumino term and its quantization},
Phys. Lett. {\bf B733} (2014) 309
[arXiv:1403.4127 [hep-th]].

\bibitem{BBTD88}
L.C.\,Biedenharn, H.W.\,Braden, P.\,Truini, H.\,van\,Dam,
{\it Relativistic wavefunctions on spinor spaces},
J. Phys. {\bf A21} (1988) 3593.

\bibitem{Sok86}
E.\,Sokatchev,
{\it Light cone harmonic superspace and
its applications}, Phys. Lett. {\bf B169} 1986) 209;
{\it Harmonic superparticle}, Class. Quant. Grav. {\bf 4} (1987) 237.

\bibitem{DKS}
F.\,Delduc, S.\,Kalitsyn, E.\,Sokatchev,
{\it Learning the ABC of light cone harmonic space},
Class. Quant. Grav. {\bf 6} (1989) 1561.

\bibitem{Jack:91}
R.\,Jackiw, V.P.\,Nair, {\it Relativistic wave equation for
anyons}, Phys. Rev. {\bf D43}, 1933-1942 (1991).

\bibitem{SV93}
D.P.\,Sorokin, D.\,Volkov,
{\it (Anti)commuting spinors and supersymmetric dynamics of
semions}, Nucl. Phys. {\bf B409} (1993) 547.

\bibitem{CorPl}
J.L.\,Cort\'es, M.S.\,Plyushchay,
{\it Anyons: Minimal and extended formulations},
Mod. Phys. Lett. {\bf A10} (1995) 409 [arXiv:hep-th/9405181];
{\it Anyons as spinning particles},
Int. J. Mod. Phys. {\bf A11} (1996) 3331 [arXiv:hep-th/9505117].

\bibitem{Vil68}
N.J.\,Vilenkin,
{\it Special functions and the theory of group representations},
American Mathematical Soc., Translations of Mathematical Monographs, Vol.\,22, 1968, 613pp.

\bibitem{Bar47}
V.\,Bargmann,
{\it Irreducible unitary representations of the Lorentz group},
Annals Math. {\bf 48} (1947) 568.

\bibitem{Ruhl}
W.\,R\"{u}hl,
{\it The Lorentz group and harmonic analysis},
W.\,A.\,Benjamin, 1970, 299pp.

\bibitem{Band}
I.A.\,Bandos, {\it Superparticle in Lorentz harmonic superspace},
Sov. J. Nucl. Phys. {\bf 51} (1990) 906.

\bibitem{DGS92}
F.\,Delduc, A.\,Galperin, E.\,Sokatchev,
{\it Lorentz harmonic (super)fields and (super)particles},
Nucl. Phys. B368 (1992) 143.

\bibitem{FedZim}
S.\,Fedoruk, V.G.\,Zima,
{\it Covariant quantization of d=4 Brink-Schwarz superparticle with Lorentz harmonics},
Theor. Math. Phys. {\bf 102} (1995) 305 [arXiv:hep-th/9409117].

\bibitem{BMSS83}
A.P.\,Balachandran, G.\,Marmo, B.S.\,Skagerstam, A.\,Stern,
{\it Gauge symmetries and fiber bundles: applications to particle dynamics},
Lect. Notes in Phys. {\bf 188} (1983) 1.

\bibitem{Schul68}
L.\,Schulman,
{\it A Path Integral for Spin},
Phys. Rev. {\bf 176} (1968) 1558.

\bibitem{Orz}
C.A.\,Orzalesi,
{\it Charges and generators of symmetry transformations in quantum field theory},
Rev. Mod. Phys. {\bf 42} (1970) 381.

\bibitem{GelGV}
I.M.\,Gelfand, M.I.\,Graev, N.J.\,Vilenkin,
{\it Generalized functions: fntegral geometry and representation theory},
New York, London: Academic Press, 1966.

\bibitem{GIOS}
A.S.\,Galperin, E.A.\,Ivanov, V.I.\,Ogievetsky, E.S.\,Sokatchev,
{\it Harmonic Superspace},
Cambridge Univ. Press, 2001.

\bibitem{Cor:53}
E-M. Corson,
{\it Introduction to tensors, spinors and relativistic wave equations},
Blackie and Son, London and Glasgow, 1957

\bibitem{Wein}
S.\,Weinberg,
{\it Feynman rules for any spin},
Phys. Rev. {\bf 133} (1964) B1318.

\bibitem{LN91}
J.\,Lukierski, A.\,Nowicki,
{\it Quaternionic six-dimensional (super)twistor formalism and composite (super)spaces},
Mod. Phys. Lett. {\bf A6} (1991) 189.

\bibitem{HLM83}
Z.\,Hasiewicz, J.\,Lukierski, P.\,Morawiec,
{\it Quaternionic six-dimensional (super)twistor formalism and composite (super)spaces},
Phys. Lett. {\bf B130} (1983) 55.

\bibitem{DuHor}
C.\,Duval, P.A.\,Horvathy,
{\it Chiral fermions as classical massless spinning particles},
[arXiv:1406.0718 [hep-th]].

\bibitem{AITG:89}
J.A.\,de\,Azc\'arraga, J.P.\,Gauntlett, J.M.\,Izquierdo, P.K.\,Townsend,
{\it Topological extensions of the supersymmetry algebra for extended objects},
Phys.\ Rev.\ Lett.\  {\bf 63}, 2443 (1989).

\bibitem{A-I:99-08}
C.\,Chryssomalakos, J.A.\,de\,Azc\'arraga, J.M.\,Izquierdo, J.C.\,P\'erez\,Bueno,
{\it The geometry of branes and extended superspaces},
Nucl. Phys. {\bf B567}, 293-330 (2000)
[hep-th/9904137];
J.A.\,de\,Azc\'arraga, J.M.\,Izquierdo,
{\it Some geometrical aspects of M-theory}, 2007 Lisbon Conf. on
{\it Geometry and Physics}, AIP Conf. Proc.  {\bf 1023}, 57-70 (2008).

\bibitem{dAL82}
J.A.\,de\,Azc\'arraga, J.\,Lukierski,
{\it Supersymmetric particles with internal symmetries and central sharges},
Phys. Lett. {\bf B113} (1982) 170;
{\it Gupta-Bleuler quantization of massive superparticle models in $D=6$, $D=8$ and $D=10$},
Phys.\ Rev. {\bf D38} (1988) 509.


\bibitem{FradVas87}
E.S.\,Fradkin, M.A.\,Vasiliev,
{\it Cubic interaction in extended theories of massless higher spin fields},
Nucl. Phys. {\bf B291} (1987) 141.

\bibitem{DaVel70}
H.\,van\,Dam, M.J.G.\,Veltman,
{\it Massive and massless Yang-Mills and gravitational fields},
Nucl. Phys. {\bf B22} (1970) 397.

\bibitem{Zakh70}
V.I.\,Zakharov,
{\it Linearized gravitation theory and the graviton mass},
JETP Lett. {\bf 12} (1970) 312.

\bibitem{Por01}
M.\,Porrati,
{\it No van Dam-Veltman-Zakharov discontinuity in AdS space},
Phys. Lett. {\bf B498} (2001) 92 [arXiv:hep-th/0011152].

\bibitem{DesWal01}
S.\,Deser, A.\,Waldron,
{\it (Dis)continuities of massless limits in spin 3/2 mediated interactions and cosmological supergravity},
Phys. Lett. {\bf B501} (2001) 134 [arXiv:hep-th/0012014].

\bibitem{OgiPolu65}
V.I.\,Ogievetsky, I.V.\,Polubarinov,
{\it Interacting field of spin 2 and the Einstein equations},
Ann. Phys. (NY) {\bf 35} (1965) 167.

\bibitem{Stuec38}
E.C.G.\,Stueckelberg,
{\it Interaction energy in electrodynamics and in the field theory of nuclear forces},
Helv. Phys. Acta {\bf 11} (1938) 225.

\bibitem{Luk65}
J.\,Lukierski,
{\it Renormalizability of higher-spin theories},
Nuovo Cimento {\bf 38} (1965) 1407.

\bibitem{Ru04}
H.\,Ruegg, M.\,Ruiz-Altaba,
{\it The Stueckelberg field},
Int. J. Mod. Phys. {\bf A19} (2004) 3265 [arXiv:hep-th/0304245].

\bibitem{Met-Stuec}
R.R.\,Metsaev,
{\it Arbitrary spin conformal fields in (A)dS},
Nucl. Phys. {\bf B885} (2014) 734 [arXiv:1404.3712 [hep-th]].

\bibitem{BeiMorS04}
N.\,Beisert, M.\,Bianchi, J.F.\,Morales, H.\,Samtleben,
{\it Higher spin symmetry and N=4 SYM},
JHEP {\bf 0407} (2004) 058 [arXiv:hep-th/0405057].

\bibitem{BiaHes05}
M.\,Bianchi, P.J.\,Heslop, F.\,Riccioni,
{\it More on La Grande Bouffe},
JHEP {\bf 0508} (2005) 088 [arXiv:hep-th/0504156].

\bibitem{Vasil13}
M.A.\,Vasiliev,
{\it Multiparticle extension of the higher-spin algebra},
Class. Quant. Grav. {\bf 30} (2013) 104006 [arXiv:1212.6071 [hep-th]].

\bibitem{GelVas13}
O.A.\,Gelfond, M.A.\,Vasiliev,
{\it Operator algebra of free conformal currents via twistors},
Nucl. Phys. {\bf B876} (2013) 871 [arXiv:1301.3123 [hep-th]].
 	
\bibitem{Vasil88}
M.A.\,Vasiliev,
{\it Extended higher spin superalgebras and their realizations in terms of quantum operators},
Fortsch. Phys. {\bf 36} (1988) 33.

\bibitem{KonsVas90}
S.E.\,Konstein, M.A.\,Vasiliev,
{\it Extended higher spin superalgebras and their massless representations},
Nucl. Phys. {\bf B331} (1990) 475.

\bibitem{Boul13}
N.\,Boulanger, D.\,Ponomarev, E.D.\,Skvortsov, M.\,Taronna,
{\it On the uniqueness of higher-spin symmetries in AdS and CFT},
Int. J. Mod. Phys. {\bf A28} (2013) 1350162 [arXiv:1305.5180 [hep-th]].

\bibitem{LeeWeiZum67}
T.D.\,Lee, S.\,Weinberg, B.\,Zumino,
{\it Algebra of fields},
Phys. Rev. Lett. {\bf 18} (1967) 1029.

\bibitem{Saku69}
J.J.\,Sakurai,
{\it Currents and Mesons},
The University of Chicago Press, Chicago and London, 1969.
 	
\bibitem{FlatoFron78}
M.\,Flato, C.\,Fronsdal,
{\it One massless particle equals two Dirac singletons: elementary particles in a curved space. 6.},
Lett. Math. Phys. {\bf 2} (1978) 421-426.


\bibitem{Sour}
J.-M.\,Souriau, {\it Structure des syst\`emes dynamiques}, Dunod (1970)
[Engl. trans. {\it Structure of dynamical systems: a symplectic view of physics},
Progress in Mathematics, vol.149. (Boston, Mass., 1997)]

\bibitem{Kunz}
H.P.\,K\"{u}nzle,
{\it Canonical dynamics of spinning particles in gravitational and electromagnetic fields},
J. Math. Phys. {\bf 13} (1972) 739.

\bibitem{PTs98}
A.\,Pashnev, M.\,Tsulaia,
{\it Description of the higher massless irreducible integer spins in the BRST approach},
Mod. Phys. Lett. {\bf A13} (1998) 1853-1864
[arXiv:hep-th/9803207].

\bibitem{BPTs01}
C.\,Burdik, A.\,Pashnev, M.\,Tsulaia,
{\it On the mixed symmetry irreducible representations of the Poincare group in the BRST approach},
Mod. Phys. Lett. {\bf A16} (2001) 731
[arXiv:hep-th/0101201].

\bibitem{BuchPT01}
I.L.\,Buchbinder, A.\,Pashnev, M.\,Tsulaia,
{\it Lagrangian formulation of the massless higher integer spin fields in the AdS background},
Phys. Lett. {\bf B523} (2001) 338
[arXiv:hep-th/0109067].

\bibitem{BuchKT15}
I.L.\,Buchbinder, V.A.\,Krykhtin, M.\,Tsulaia,
{\it Lagrangian formulation of massive fermionic higher spin fields on a constant electromagnetic background},
[arXiv:1501.03278 [hep-th]].


\end{thebibliography}
\end{document}